\expandafter\edef\csname hypers\string @fe\endcsname{\catcode
                                             `\noexpand @=\the\catcode`\@}%
\catcode`\@=11
%
%
\ifx\hyper@utoprocess\hyper@ndefined
\else
 \expandafter\hyper@utoprocess\fi
\ifx\hyperd@ne\hyper@ndefined
 \global\let\hyperd@ne=\relax
\else
 \hypers@fe
 \errhelp{hyperbasics.tex needs to be included only once outside
          of any {...} or \begingroup...\endgroup. You have tried to
          include it more than once. If the previous include was indeed
          outside any groupings, continue and all will be well.}%
 \errmessage{Input this file only once!}%
 \expandafter \fi
%
%
\def\hyperv@rsion{12}%
%
%
\newread\hyperf@le
\def\hyperf@lename{\jobname.hrf}%
\immediate\openin\hyperf@le\hyperf@lename\relax
\ifeof\hyperf@le\relax
 \immediate\closein\hyperf@le\relax
\else
 \immediate\closein\hyperf@le\relax
 \input \hyperf@lename
\fi
%
%
\newwrite\hyperf@le
\immediate\openout\hyperf@le\hyperf@lename
%
%
\newtoks\hypert@ks
%
%
\edef\hypert@mp{\catcode`\noexpand\%=\the\catcode`\%}
\catcode`\%=12
\def\hyperp@rcent{
\hypert@mp
\edef\hypert@mp{\catcode`\noexpand\#=\the\catcode`\#}%
\catcode`\#=12
\def\hyperh@sh{#}%
\hypert@mp
\let\hypert@mp=\relax
\let\hyper@nd=\relax
\def\hyperbl@nk{ }
\def\hyperstr@pquote#1"#2\hyper@nd{
   #1
   \ifx\hyper@nd#2\hyper@nd
   \else\hyperp@rcent22\hyperstr@pquote#2\hyper@nd\fi}%
\def\hyperstr@pblank#1 #2\hyper@nd{
   #1
   \ifx\hyper@nd#2\hyper@nd
   \else\hyperp@rcent20\hyperstr@pblank#2\hyper@nd\fi}
\long\def\hyper@nchor#1#2{\edef\hyperm@cro{html:<A #1>}%
                          \special\expandafter{\hyperm@cro}%
                          {#2}}%
\def\hyper@atm@ning#1->#2\hyper@nd{#2}
\def\hyperlink{\protect\hyperlink@}
\def\hyperlink@{{\catcode\expandafter`\noexpand\#=12 
                 \catcode\expandafter`\noexpand\%=12 
                 \catcode\expandafter`\noexpand\~=12
                \expandafter}\hyperlink@@}
\def\hyperlink@@#1{\protect\hyperlink@@@{#1}}
\def\hyperlink@@@#1{\edef\hypert@mp{#1}%
               \edef\hypert@mp{\expandafter\hyper@atm@ning\meaning\hypert@mp
                               \hyper@nd}%
               \edef\hypert@mp{\expandafter\hyperstr@pquote%
                               \hypert@mp"\hyper@nd}%
               \edef\hypert@mp{\expandafter\expandafter\expandafter
                               \hyperstr@pblank\expandafter%
                               \hypert@mp\hyperbl@nk\hyper@nd}%
               \hyper@nchor{href=\expandafter"\hypert@mp"}}%
\def\hypertarget#1{\edef\hypert@mp{#1}%
               \edef\hypert@mp{\expandafter\hyper@atm@ning\meaning\hypert@mp
                               \hyper@nd}%
               \edef\hypert@mp{\expandafter\hyperstr@pquote%
                               \hypert@mp"\hyper@nd}%
               \edef\hypert@mp{\expandafter\expandafter\expandafter
                               \hyperstr@pblank\expandafter%
                               \hypert@mp\hyperbl@nk\hyper@nd}%
               \hyper@nchor{name=\expandafter"\hypert@mp"}}%
\def\hyperref{\afterassignment\hyperr@f\let\hyperp@ram}
\def\hyperr@f{\ifx\hyperp@ram{\iffalse}\fi
               \expandafter\expandafter\expandafter\hyperr@@
               \expandafter{%
              \else
               \iffalse}\fi
               \ifx\hyperp@ram\hyper@ndefined
                 \message{Undefined reference}%
                 \def\hyperp@r@m{{}{undefined}{}}%
               \else
                 \edef\hyperp@r@m{\hyperp@ram}%
               \fi
               \expandafter\expandafter\expandafter\hyperr@@
               \expandafter\hyperp@r@m
              \fi}%
\def\hyperr@@#1#2#3{\ifx\hyper@ndefined#1\hyper@ndefined
                    \hypert@ks\expandafter{\hyperh@sh#2.#3}%
                    \else
                     \ifx\hyper@ndefined#2#3\hyper@ndefined
                      \hypert@ks{#1}%
                     \else
                      \def\hypert@mp{#1}%
                      \hypert@ks\expandafter\expandafter\expandafter
                      {\expandafter\hypert@mp\hyperh@sh#2.#3}%
                     \fi
                    \fi
                    \expandafter\hyperlink\expandafter{\the\hypert@ks}}%
\def\hyperdef#1#2#3{{\escapechar=`\\\relax
                     \edef\hyper@t@mp@{\hyperstr@pquote#2.#3"\hyper@nd}%
                     \expandafter\ifx\csname hyperd@\meaning\hyper@t@mp@
                     \endcsname
                     \relax
                     \expandafter\gdef\csname hyperd@\meaning\hyper@t@mp@
                     \endcsname{}%
                     \gdef#1{{}{\hyperstr@pquote#2"\hyper@nd}%
                               {\hyperstr@pquote#3"\hyper@nd}}%
                     \immediate\write\hyperf@le{\def\noexpand#1{{}%
                        {\noexpand\hyperstr@pquote#2"\noexpand\hyper@nd}%
                        {\noexpand\hyperstr@pquote#3"\noexpand\hyper@nd}}}%
                     \xdef\hyper@t@mp@{\global\let\noexpand\hyper@t@mp@=\relax
                                       \noexpand\hypertarget{\hyper@t@mp@}}%
                     \global\hypert@ks={\hyper@t@mp@}%
                     \else
                     \message\expandafter{\expandafter'\hyper@t@mp@' duplicate}%
		     \def\hyper@@tmp@{\hyperdef{#1}{#2}}%
		     \edef\hyper@@tmp@@{{#3@}}%
                     \global\let\hyper@t@mp@=\relax
                     \global\hypert@ks=\expandafter\expandafter
		     \expandafter{\expandafter\hyper@@tmp@\hyper@@tmp@@}%
                     \fi}\the\hypert@ks}%

\def\hyper@nique#1#2#3#4{{\escapechar=`\\\relax
                     \edef\hyper@t@mp@{\hyperstr@pquote#2.#3"\hyper@nd}%
                     \expandafter\ifx\csname hyperd@\meaning\hyper@t@mp@
                     \endcsname
                     \relax
                     \gdef#1{{}{\hyperstr@pquote#2"\hyper@nd}%
                               {\hyperstr@pquote#3"\hyper@nd}}%
                     \global\let\hyper@t@mp@=\relax
                     #4%
                     \else
		     \def\hyper@@tmp@{\hyper@nique{#1}{#2}}%
		     \edef\hyper@@tmp@@{{#3@}}%
                     \global\let\hyper@t@mp@=\relax
   		     \expandafter\hyper@@tmp@\hyper@@tmp@@{#4}%
                     \fi}}%

\ifx\protect\hyper@ndefined\let\protect=\relax\fi
\let\hyper@@@@=\relax
\def\hyper@@{\let\hyper@@@=\relax}%
\hyper@@
\def\hyper@{\relax\let\hyper@@@\noexpand\hyper@\noexpand}%
\def\hyperpr@ref{\hyper@@\hyperref}
\def\hyperpr@link{\hyper@@\hyperlink}
\def\hyperpr@def{\hyper@@\hyperdef}
\let\hyper@marker=\relax
\def\hyper@@tokenize#1\hyper@marker{#1}
\def\hyper@tokenize{\expandafter\endgroup\hyper@@tokenize}
\def\hyperpr@tokenize{\hyper@@\hyper@tokenize}

\edef\href{\begingroup\catcode`\string @=11
            \hyper@\hyperpr@tokenize\hyper@\hyperpr@link
            \hyper@marker}
\let\hname\hypertarget
\def\allowoncemore{\def\hyper@utoprocess{\let\hyper@utoprocess=\hyper@ndefined
                                         \hypers@fe }}%
%
%
\hypers@fe
 

\input amssym.tex 

\def\unredoffs{}
\tolerance=1000\hfuzz=2pt\def\fontflag{cm}
\catcode`\@=11 
\ifx\hyperdef\UNd@FiNeD\def\hyperdef#1#2#3#4{#4}\def\hyperref#1#2#3#4{#4}\def\href#1#2{#2}\fi
\magnification=1200\unredoffs\baselineskip=16pt plus 2pt minus 1pt
\def\Title#1#2{\nopagenumbers\rightline{#1}%
\vskip 1in\centerline{#2}\vskip .5in\pageno=0}
\def\Date#1{\vfill\leftline{#1}\tenpoint\supereject%
\footline={\hss\tenrm\hyperdef\hypernoname{page}\folio\folio\hss}}%

{\count255=\time\divide\count255 by 60 \xdef\hourmin{\number\count255}
 \multiply\count255 by-60\advance\count255 by\time
 \xdef\hourmin{\hourmin:\ifnum\count255<10 0\fi\the\count255}
}
\def\date{\number\day.\number\month.\number\year\ at \hourmin}

\def\draft{\message{ DRAFTMODE }\def\draftdate{{\rm Preliminary draft: \date}}%
\headline={\hfil\draftdate}\writelabels\baselineskip=20pt plus 2pt minus 2pt}
\def\nolabels{\def\wrlabeL##1{}\def\eqlabeL##1{}\def\reflabeL##1{}}
\def\writelabels{\def\wrlabeL##1{\leavevmode\vadjust{\rlap{\smash%
{\line{{\escapechar=` \hfill\rlap{\sevenrm\hskip.03in\string##1}}}}}}}%
\def\eqlabeL##1{{\escapechar-1\rlap{\sevenrm\hskip.05in\string##1}}}%
\def\reflabeL##1{\noexpand\llap{\noexpand\sevenrm\string\string\string##1}}}
\nolabels

\global\newcount\secno \global\secno=0
\global\newcount\meqno \global\meqno=1
\def\s@csym{}

\def\newsec#1\par{\global\advance\secno by1%
{\toks0{#1}\message{(\the\secno. \the\toks0)}}%
\global\subsecno=0\eqnres@t\let\s@csym\secsym\xdef\secn@m{\the\secno}\noindent
{\bf\hyperdef\hypernoname{section}{\the\secno}{\the\secno.} #1}%
\writetoca{{\string\hyperref{}{section}{\the\secno}{\bf \the\secno\quad}} {\bf #1}}\par%
\nobreak\medskip\nobreak\noindent\ignorespaces}
\def\eqnres@t{\xdef\secsym{\the\secno.}\global\meqno=1\bigbreak\bigskip}
\def\sequentialequations{\def\eqnres@t{\bigbreak}}\xdef\secsym{}

\global\newcount\subsecno \global\subsecno=0
\def\subsec#1\par{\global\advance\subsecno by1%
{\toks0{#1}\message{(\s@csym\the\subsecno. \the\toks0)}}%
\global\subsubsecno=0%
\ifnum\lastpenalty>9000\else\bigbreak\fi
\noindent{\it\hyperdef\hypernoname{subsection}{\secn@m.\the\subsecno}%
{\secn@m.\the\subsecno.} #1}\writetoca{\string\hskip1.45cm
{\string\hyperref{}{subsection}{\secn@m.\the\subsecno}{\secn@m.\the\subsecno.}}
{#1}}\par\nobreak\medskip\nobreak\noindent\ignorespaces}

\global\newcount\subsubsecno \global\subsubsecno=0
\def\subsubsec#1\par{\global\advance\subsubsecno by1%
{\toks0{#1}\message{(\secn@m.\the\subsecno.\the\subsubsecno. \the\toks0)}}%
\global\subsubsubsecno=0%
\ifnum\lastpenalty>9000\else\bigbreak\fi
\noindent{\it\hyperdef\hypernoname{subsubsection}{\secn@m.\the\subsecno\the\subsubsecno}%
{\secn@m.\the\subsecno.\the\subsubsecno.} #1}
\par\nobreak\medskip\nobreak\noindent\ignorespaces}

\global\newcount\subsubsubsecno \global\subsubsubsecno=0
\def\subsubsubsec#1\par{\global\advance\subsubsubsecno by1%
{\toks0{#1}\message{(\secn@m.\the\subsecno.\the\subsubsecno.\the\subsubsubsecno \the\toks0)}}%
\ifnum\lastpenalty>9000\else\bigbreak\fi
\noindent{\it\hyperdef\hypernoname{subsubsection}{\secn@m.\the\subsecno\the\subsubsecno\the\subsubsubsecno}%
{\secn@m.\the\subsecno.\the\subsubsecno.\the\subsubsubsecno.} #1}%
\par\nobreak\medskip\nobreak\noindent\ignorespaces}


\def\newnewsec#1#2\par{\global\advance\secno by1%
{\toks0{#2}\message{(\secn@m. \the\toks0)}}%
\global\subsecno=0\global\subsubsecno=0\eqnres@t\let\s@csym\secsym\xdef\secn@m{\the\secno}\noindent
\ifnum\lastpenalty>9000\else\bigbreak\fi
\noindent{\bf\hyperdef\hypernoname{section}{\secn@m}{\secn@m.} #2}%
\writetoca{{\string\hyperref{}{section}{\the\secno}{\bf \the\secno\quad}} {\bf #2}}
\DefWarn#1%
\xdef#1{\noexpand\hyperref{}{section}{\the\secno}%
{\the\secno}}\writedef{#1\leftbracket#1}\wrlabeL{#1=#1}%
\par\nobreak\medskip\nobreak\noindent\ignorespaces}

\def\newsubsec#1#2\par{\global\advance\subsecno by1%
{\toks0{#2}\message{(\secn@m.\the\subsecno. \the\toks0)}}%
\global\subsubsecno=0%
\ifnum\lastpenalty>9000\else\bigbreak\fi
\noindent{\it\hyperdef\hypernoname{subsection}{\secn@m.\the\subsecno}%
{\secn@m.\the\subsecno.} #2}
\DefWarn#1%
\xdef#1{\noexpand\hyperref{}{subsection}{\secn@m.\the\subsecno}%
{\secn@m.\the\subsecno}}\writedef{#1\leftbracket#1}\wrlabeL{#1=#1}%
\writetoca{\string\hskip1.45cm
{\string\hyperref{}{subsection}{\secn@m.\the\subsecno}{\secn@m.\the\subsecno.}}
{#2}}%
\par\nobreak\medskip\nobreak\noindent\ignorespaces}

\def\newsubsecstar#1#2\par{\global\advance\subsecno by1%
{\toks0{#2}\message{(\secn@m.\the\subsecno. \the\toks0)}}%
\global\subsubsecno=0%
\ifnum\lastpenalty>9000\else\bigbreak\fi
\noindent{\it\hyperdef\hypernoname{subsection}{\secn@m.\the\subsecno}%
{\secn@m.\the\subsecno.} #2}
\DefWarn#1%
\xdef#1{\noexpand\hyperref{}{subsection}{\secn@m.\the\subsecno}%
{\secn@m.\the\subsecno}}\writedef{#1\leftbracket#1}\wrlabeL{#1=#1}%
\par\nobreak\medskip\nobreak\noindent\ignorespaces}

\def\newsubsubsec#1#2\par{\global\advance\subsubsecno by1%
{\toks0{#2}\message{(\secn@m.\the\subsecno.\the\subsubsecno. \the\toks0)}}%
\global\subsubsubsecno=0%
\ifnum\lastpenalty>9000\else\bigbreak\fi
\noindent{\it\hyperdef\hypernoname{subsubsection}{\secn@m.\the\subsecno.\the\subsubsecno}%
{\secn@m.\the\subsecno.\the\subsubsecno.} #2}
\DefWarn#1%
\xdef#1{\noexpand\hyperref{}{subsubsection}{\secn@m.\the\subsecno.\the\subsubsecno}%
{\secn@m.\the\subsecno.\the\subsubsecno}}\writedef{#1\leftbracket#1}\wrlabeL{#1=#1}%
\par\nobreak\medskip\nobreak\noindent\ignorespaces}

\def\newsubsubsubsec#1#2\par{\global\advance\subsubsubsecno by1%
{\toks0{#2}\message{(\secn@m.\the\subsecno.\the\subsubsecno.\the\subsubsubsecno \the\toks0)}}%
\ifnum\lastpenalty>9000\else\bigbreak\fi
\noindent{\it\hyperdef\hypernoname{subsubsection}{\secn@m.\the\subsecno\the\subsubsecno\the\subsubsubsecno}%
{\secn@m.\the\subsecno.\the\subsubsecno.\the\subsubsubsecno.} #2}
\DefWarn#1%
\xdef#1{\noexpand\hyperref{}{subsubsubsection}{\secn@m.\the\subsecno.\the\subsubsecno.\the\subsubsubsecno}%
{\secn@m.\the\subsecno.\the\subsubsecno.\the\subsubsubsecno}}\writedef{#1\leftbracket#1}\wrlabeL{#1=#1}%
\par\nobreak\medskip\nobreak\noindent\ignorespaces}

\def\appendix#1#2{\global\meqno=1\global\subsecno=0\global\subsubsecno=0\xdef\secsym{\hbox{#1.}}%
\bigbreak\bigskip\noindent{\bf Appendix \hyperdef\hypernoname{appendix}{#1}%
{#1.} #2}{\toks0{(#1. #2)}\message{\the\toks0}}%
\xdef\s@csym{#1.}\xdef\secn@m{#1}%
\writetoca{{\string\hyperref{}{appendix}{#1}{\bf {#1}\quad}} {\bf #2}}%
\par\nobreak\medskip\nobreak}

%
\def\checkm@de#1#2{\ifmmode{\def\f@rst##1{##1}\hyperdef\hypernoname{equation}%
{#1}{#2}}\else\hyperref{}{equation}{#1}{#2}\fi}
\def\eqnn#1{\DefWarn#1\xdef #1{(\noexpand\relax\noexpand\checkm@de%
{\s@csym\the\meqno}{\secsym\the\meqno})}%
\wrlabeL#1\writedef{#1\leftbracket#1}\global\advance\meqno by1}
\def\f@rst#1{\c@t#1a\em@ark}\def\c@t#1#2\em@ark{#1}
\def\eqna#1{\DefWarn#1\wrlabeL{#1$\{\}$}%
\xdef #1##1{(\noexpand\relax\noexpand\checkm@de%
{\s@csym\the\meqno\noexpand\f@rst{##1}1}{\hbox{$\secsym\the\meqno##1$}})}
\writedef{#1\numbersign1\leftbracket#1{\numbersign1}}\global\advance\meqno by1}
\def\eqn#1#2{\DefWarn#1%
\xdef #1{(\noexpand\hyperref{}{equation}{\s@csym\the\meqno}%
{\secsym\the\meqno})}$$#2\eqno(\hyperdef\hypernoname{equation}%
{\s@csym\the\meqno}{\secsym\the\meqno})\eqlabeL#1$$%
\writedef{#1\leftbracket#1}\global\advance\meqno by1}
\def\xeqn{\expandafter\xe@n}\def\xe@n(#1){#1}
\def\xeqna#1{\expandafter\xe@n#1}
\def\eqns#1{(\e@ns #1{\hbox{}})}
\def\e@ns#1{\ifx\UNd@FiNeD#1\message{eqnlabel \string#1 is undefined.}%
\xdef#1{(?.?)}\fi{\let\hyperref=\relax\xdef\next{#1}}%
\ifx\next\em@rk\def\next{}\else%
\ifx\next#1\xeqn#1\else\def\n@xt{#1}\ifx\n@xt\next#1\else\xeqna#1\fi
\fi\let\next=\e@ns\fi\next}
\def\etag#1{\eqnn#1\eqno#1}\def\etaga#1{\eqna#1\eqno#1}
\def\DefWarn#1{}
%
\newskip\footskip\footskip14pt plus 1pt minus 1pt 
\def\footnotefont{\ninepoint}\def\f@t#1{\footnotefont #1\@foot}
\def\f@@t{\baselineskip\footskip\bgroup\footnotefont\aftergroup\@foot\let\next}
\setbox\strutbox=\hbox{\vrule height9.5pt depth4.5pt width0pt}
\global\newcount\ftno \global\ftno=0
\def\foot{\global\advance\ftno by1\def\foot@rg{\hyperref{}{footnote}%
{\the\ftno}{\the\ftno}\xdef\foot@rg{\noexpand\hyperdef\noexpand\hypernoname%
{footnote}{\the\ftno}{\the\ftno}}}\footnote{$^{\foot@rg}$}}
%
%
%
\global\newcount\refno \global\refno=1
\newwrite\rfile
\def\ref{[\hyperref{}{reference}{\the\refno}{\the\refno}]\nref}
\def\nref#1{\DefWarn#1%
\xdef#1{[\noexpand\hyperref{}{reference}{\the\refno}{\the\refno}]}%
\writedef{#1\leftbracket#1}%
\ifnum\refno=1\immediate\openout\rfile=\jobname.refs\fi
\chardef\wfile=\rfile\immediate\write\rfile{\noexpand\item{[\noexpand\hyperdef%
\noexpand\hypernoname{reference}{\the\refno}{\the\refno}]\ }%
\reflabeL{#1\hskip.31in}\pctsign}\global\advance\refno by1\findarg}
\def\findarg#1#{\begingroup\obeylines\newlinechar=`\^^M\pass@rg}
{\obeylines\gdef\pass@rg#1{\writ@line\relax #1^^M\hbox{}^^M}%
\gdef\writ@line#1^^M{\expandafter\toks0\expandafter{\striprel@x #1}%
\edef\next{\the\toks0}\ifx\next\em@rk\let\next=\endgroup\else\ifx\next\empty%
\else\immediate\write\wfile{\the\toks0}\fi\let\next=\writ@line\fi\next\relax}}
\def\striprel@x#1{} \def\em@rk{\hbox{}}
\def\lref{\begingroup\obeylines\lr@f}
\def\lr@f#1#2{\DefWarn#1\gdef#1{\let#1=\UNd@FiNeD\ref#1{#2}}\endgroup\unskip}
\def\semi{;\hfil\break}
\def\addref#1{\immediate\write\rfile{\noexpand\item{}#1}} 
\def\listrefs{\vfill\supereject\immediate\closeout\rfile\writestoppt
\baselineskip=\footskip\centerline{{\bf References}}\bigskip{\parindent=20pt%
\frenchspacing\escapechar=` \input \jobname.refs\vfill\eject}\nonfrenchspacing}
\def\startrefs#1{\immediate\openout\rfile=\jobname.refs\refno=#1}
\def\xref{\expandafter\xr@f}\def\xr@f[#1]{#1}
\def\refs#1{\count255=1[\r@fs #1{\hbox{}}]}
\def\r@fs#1{\ifx\UNd@FiNeD#1\message{reflabel \string#1 is undefined.}%
\nref#1{need to supply reference \string#1.}\fi%
\vphantom{\hphantom{#1}}{\let\hyperref=\relax\xdef\next{#1}}%
\ifx\next\em@rk\def\next{}%
\else\ifx\next#1\ifodd\count255\relax\xref#1\count255=0\fi%
\else#1\count255=1\fi\let\next=\r@fs\fi\next}
\def\figures{\centerline{{\bf Figure Captions}}\medskip\parindent=40pt%
\def\fig##1##2{\medskip\item{Fig.~\hyperdef\hypernoname{figure}{##1}{##1}.  }%
##2}}
%
\newwrite\ffile\global\newcount\figno \global\figno=1
\def\fig{fig.~\hyperref{}{figure}{\the\figno}{\the\figno}\nfig}
\def\nfig#1{\DefWarn#1%
\xdef#1{fig.~\noexpand\hyperref{}{figure}{\the\figno}{\the\figno}}%
\writedef{#1\leftbracket fig.\noexpand~\xfig#1}%
\ifnum\figno=1\immediate\openout\ffile=\jobname.figs\fi\chardef\wfile=\ffile%
{\let\hyperref=\relax
\immediate\write\ffile{\noexpand\medskip\noexpand\item{Fig.\ %
\noexpand\hyperdef\noexpand\hypernoname{figure}{\the\figno}{\the\figno}. }
\reflabeL{#1\hskip.55in}\pctsign}}\global\advance\figno by1\findarg}
\def\xfig{\expandafter\xf@g}\def\xf@g fig.\penalty\@M\ {}
\def\figs#1{figs.~\f@gs #1{\hbox{}}}
\def\f@gs#1{{\let\hyperref=\relax\xdef\next{#1}}\ifx\next\em@rk\def\next{}\else
\ifx\next#1\xfig #1\else#1\fi\let\next=\f@gs\fi\next}
%
\def\figin{\epsfcheck\figin}\def\figins{\epsfcheck\figins}
\def\epsfcheck{\ifx\epsfbox\UnDeFiNeD
\message{(NO epsf.tex, FIGURES WILL BE IGNORED)}
\gdef\figin##1{\vskip2in}\gdef\figins##1{\hskip.5in}
\else\message{(FIGURES WILL BE INCLUDED)}%
\gdef\figin##1{##1}\gdef\figins##1{##1}\fi}
\def\figinsert{\goodbreak\topinsert}
\def\ifig#1#2#3{\DefWarn#1\xdef#1{fig.~\the\figno}
\writedef{#1\leftbracket fig.\noexpand~\the\figno}%
\figinsert\figin{\centerline{#3}}
\smallskip
\leftskip=0pt \rightskip=0pt
\baselineskip12pt\noindent
{{\bf Fig.~\the\figno}\ \ninepoint #2}
\medskip
\global\advance\figno by1\par\endinsert}
\newwrite\lfile
{\escapechar-1\xdef\pctsign{\string\%}\xdef\leftbracket{\string\{}
\xdef\rightbracket{\string\}}\xdef\numbersign{\string\#}}
\def\writedefs{\immediate\openout\lfile=label.defs \def\writedef##1{%
{\let\hyperref=\relax\let\hyperdef=\relax\let\hypernoname=\relax
 \immediate\write\lfile{\string\checkdef\string##1\rightbracket}}}}%
\def\writestop{\def\writestoppt{\immediate\write\lfile{\string\pageno
 \the\pageno\string\startrefs\leftbracket\the\refno\rightbracket
 \string\def\string\secsym\leftbracket\secsym\rightbracket
 \string\secno\the\secno\string\meqno\the\meqno}\immediate\closeout\lfile}}
\def\writestoppt{}\def\writedef#1{}

\def\seclab#1\par{\DefWarn#1%
\xdef #1{\noexpand\hyperref{}{section}{\the\secno}{\the\secno}}%
\writedef{#1\leftbracket#1}\wrlabeL{#1=#1}\par%
\nobreak\medskip\nobreak\noindent\ignorespaces}
\def\subseclab#1\par{\DefWarn#1%
\xdef #1{\noexpand\hyperref{}{subsection}{\the\secno.\the\subsecno}%
{\the\secno.\the\subsecno}}\writedef{#1\leftbracket#1}\wrlabeL{#1=#1}\par%
\nobreak\medskip\nobreak\noindent\ignorespaces}
\def\subsubseclab#1\par{\DefWarn#1%
\xdef#1{\noexpand\hyperref{}{subsubsection}{\the\secno.\the\subsecno.\the\subsubsecno}%
{\the\secno.\the\subsecno.\the\subsubsecno}}\writedef{#1\leftbracket#1}\wrlabeL{#1=#1}\par%
\nobreak\medskip\nobreak\noindent\ignorespaces}
\def\applab#1\par{\DefWarn#1%
\xdef#1{\noexpand\hyperref{}{appendix}{\secn@m}{\secn@m}}%
\writedef{#1\leftbracket#1}\wrlabeL{#1=#1}%
\par\nobreak\medskip\nobreak\noindent\ignorespaces}
\def\appsublab#1{\DefWarn#1%
\xdef #1{\noexpand\hyperref{}{appendix}{\secn@m.\the\subsecno}{\secn@m.\the\subsecno}}%
\writedef{#1\leftbracket#1}\wrlabeL{#1=#1}}
\newwrite\tfile \def\writetoca#1{}
\def\leaderfill{\leaders\hbox to 1em{\hss.\hss}\hfill}
\def\writetoc{\immediate\openout\tfile=\jobname.toc
   \def\writetoca##1{{\edef\next{\write\tfile{\noindent ##1
   \string\leaderfill{
   \string\hyperref{}{page}{\noexpand\number\pageno}%
   {\noexpand\number\pageno}} \par}}\next}}
}
\newread\ch@ckfile
\def\listtoc{\immediate\closeout\tfile\immediate\openin\ch@ckfile=\jobname.toc
\ifeof\ch@ckfile\message{no file \jobname.toc, no table of contents this pass}%
\else\closein\ch@ckfile\centerline{\bf Contents}\nobreak\medskip%
{\baselineskip=15.5pt\footnotefont\parskip=0pt\catcode`\@=11\input\jobname.toc
\catcode`\@=12\bigbreak\bigskip}\fi}
\catcode`\@=12 
\font\authorfont=cmcsc10
\def\tenpoint{\def\rm{\fam0\tenrm}
\textfont0=\tenrm \scriptfont0=\sevenrm \scriptscriptfont0=\fiverm
\textfont1=\teni  \scriptfont1=\seveni  \scriptscriptfont1=\fivei
\textfont2=\tensy \scriptfont2=\sevensy \scriptscriptfont2=\fivesy
\textfont\itfam=\tenit \def\it{\fam\itfam\tenit}\def\footnotefont{\ninepoint}%
\textfont\bffam=\tenbf \def\bf{\fam\bffam\tenbf}\def\sl{\fam\slfam\tensl}\rm}
\font\ninerm=cmr9 \font\sixrm=cmr6 \font\ninei=cmmi9 \font\sixi=cmmi6
\font\ninesy=cmsy9 \font\sixsy=cmsy6 \font\ninebf=cmbx9
\font\nineit=cmti9 \font\ninesl=cmsl9 \skewchar\ninei='177
\skewchar\sixi='177 \skewchar\ninesy='60 \skewchar\sixsy='60
\def\ninepoint{\def\rm{\fam0\ninerm}
\textfont0=\ninerm \scriptfont0=\sixrm \scriptscriptfont0=\fiverm
\textfont1=\ninei \scriptfont1=\sixi \scriptscriptfont1=\fivei
\textfont2=\ninesy \scriptfont2=\sixsy \scriptscriptfont2=\fivesy
\textfont\itfam=\ninei \def\it{\fam\itfam\nineit}\def\sl{\fam\slfam\ninesl}%
\textfont\bffam=\ninebf \def\bf{\fam\bffam\ninebf}\rm}

\font\eightrm=cmr8
\font\eighti=cmmi8
\font\eightsy=cmsy8
\font\eightbf=cmbx8
\font\eighttt=cmtt8
\font\eightit=cmti8
\font\eightsl=cmsl8
\font\sixbf=cmbx6

\newskip\ttglue
\def\eightpoint{\def\rm{\fam0\eightrm}
  \textfont0=\eightrm \scriptfont0=\sixrm \scriptscriptfont0=\fiverm
  \textfont1=\eighti  \scriptfont1=\sixi  \scriptscriptfont1=\fivei
  \textfont2=\eightsy \scriptfont2=\sixsy \scriptscriptfont2=\fivesy
  \textfont3=\tenex   \scriptfont3=\tenex \scriptscriptfont3=\tenex
  \textfont\itfam=\eightit  \def\it{\fam\itfam\eightit}%
  \textfont\slfam=\eightsl  \def\sl{\fam\slfam\eightsl}%
  \textfont\ttfam=\eighttt  \def\tt{\fam\ttfam\eighttt}%
  \textfont\bffam=\eightbf  \scriptfont\bffam=\sixbf
   \scriptscriptfont\bffam=\fivebf  \def\bf{\fam\bffam\eightbf}%
  \tt \ttglue=.5em plus.25em minus.15em
  \normalbaselineskip=9pt
  \setbox\strutbox=\hbox{\vrule height7pt depth2pt width0pt}%
  \let\sc=\sixrm  \let\big=\eightbig  \normalbaselines\rm}
\hyphenation{anom-aly anom-alies coun-ter-term coun-ter-terms}

\def\tikzcaption#1#2{\DefWarn#1\xdef#1{Fig.~\the\figno}
\writedef{#1\leftbracket Fig.\noexpand~\the\figno}%
{
\smallskip
\leftskip=20pt \rightskip=20pt \baselineskip12pt\noindent
{{\bf Fig.~\the\figno}\ \ninepoint #2}
\bigskip
\global\advance\figno by1 \par}}

\def\ntoalpha#1{%
\ifcase#1%
@%
\or A\or B\or C\or D\or E\or F\or G\or H\or I\or J\or K\or L\or M%
\fi
}

\global\newcount\appno \global\appno=1
\def\applab#1{\xdef #1{\ntoalpha{\appno}}\writedef{#1\leftbracket#1}\wrlabeL{#1=#1}
\global\advance\appno by1}

\def\preprint#1 #2\par{\rightline{\vbox{\baselineskip12pt\hbox{#1}\hbox{#2}}}\vskip2cm}
%
\def\title#1\par{\centerline{\bf #1}\nopagenumbers\pageno=0}
\def\author#1\par{\bigskip\bigskip\centerline{#1}}

\newcount\addressno

\def\email#1#2{
\footnote{\null}{\kern-\parindent \llap{$^#1$\hskip1pt}email: #2}}

\def\startcenter{%
  \par
  \begingroup
  \leftskip=0pt plus 1fil
  \rightskip=\leftskip
  \parindent=0pt
  \parfillskip=0pt
}
\def\stopcenter{\endgroup}

\def\address{\bigskip%
  \ifnum\the\addressno=0\else\stopcenter\endgroup\fi
  \advance\addressno by 1%
  \begingroup
  \startcenter
  \it
  \obeylines
  \addressAux
}
\def\addressAux#1{#1}

\def\abstract{\stopcenter\endgroup\bigskip\bigskip\noindent}

\def\Dsl{\,\raise.15ex\hbox{/}\mkern-13.5mu D} 
\def\dsl{\raise.15ex\hbox{/}\kern-.57em\partial}
\def\tr{{\rm tr}} \def\Tr{{\rm Tr}}
\def\boxeqn#1{\vcenter{\vbox{\hrule\hbox{\vrule\kern3pt\vbox{\kern3pt
	\hbox{${\displaystyle #1}$}\kern3pt}\kern3pt\vrule}\hrule}}}
\def\grad#1{\,\nabla\!_{{#1}}\,}
\def\gradgrad#1#2{\,\nabla\!_{{#1}}\nabla\!_{{#2}}\,}
\def\lform{\hbox{$\sqcup$}\llap{\hbox{$\sqcap$}}}
\def\lie{\hbox{\it\$}} 

\def\ap{{\alpha^{\prime}}}
\def\halfap#1{\Big({\ap\over 2}\Big)^{\mkern-4mu #1}}
\def\a{\alpha}
\def\b{{\beta}}
\def\g{{\gamma}}
\def\d{{\delta}}
\def\e{{\epsilon}}
\def\l{\lambda}
\def\k{{\kappa}}
\def\s{{\sigma}}
\def\t{{\theta}}
\def\om{{\omega}}
\def\lb{{\overline\lambda}}
\def\llb{(\l\lb)}
\def\wb{{\overline w}}
\def\half{{1\over 2}}
\def\p{{\partial}}
\def\pb{{\overline\partial}}
\def\tb{{\overline\theta}}
\def\bar{\overline}
\def\({\left(}
\def\){\right)}
\def\dz{{\rm d}z}

\def\cA{{\cal A}}
\def\cF{{\cal F}}
\def\cJ{{\cal J}}
\def\cK{{\cal K}}
\def\cI{{\cal I}}
\def\cV{{\cal V}}
\def\cW{{\cal W}}
\def\cY{{\cal Y}}
\def\cZ{{\cal Z}}

\def\bV{{\Bbb V}}
\def\bA{{\Bbb A}}
\def\bW{{\Bbb W}}
\def\bF{{\Bbb F}}

\def\Box{\square}
\def\AYM{A^{\rm SYM}}

\def\psum{\mathop{\sum\nolimits'}}

\def\len#1{{%
\def\Dlen{\left|\mkern-1mu #1\mkern -0.5mu\right|}%
\def\Sslen{\left|\mkern-1.3mu #1\mkern -1.3mu\right|}%
\def\SSlen{\left|\mkern-2.8mu #1\mkern-1.3mu\right|}%
\mathchoice{\Dlen}{\Dlen}{\Sslen}{\SSlen}}}

\def\perm#1{{\rm perm}#1}
\def\eikx{{\bigl\langle \prod_{j=1}^4 {\rm e}^{i k^j\cdot x^j}\bigr\rangle}}
\def\ImOmega{\Im\Omega}
\def\lalb#1{(\l\g^{#1}\lb)}
\def\Im{\mathop{{\rm Im}}} 
\def\sfrac#1/#2{\kern.1em\raise.5ex\hbox{\the\scriptfont0 #1}%
\kern-.1em/\kern-.15em\lower.25ex\hbox{\the\scriptfont0 #2}}

\font\tenshuffle=shuffle10 \font\sevenshuffle=shuffle7 \font\fiveshuffle=shuffle7 at 5pt
\def\shuffle{{%
\def\Dshuffle{\mathbin{\hbox{\tenshuffle\char'001}}}%
\def\Sshuffle{\mathbin{\hbox{\sevenshuffle\char'001}}}%
\def\SSshuffle{\mathbin{\hbox{\fiveshuffle\char'001}}}%
\mathchoice{\Dshuffle}{\Dshuffle}{\Sshuffle}{\SSshuffle}}}

\font\tenwedge=stmary10 \font\sevenwedge=stmary7 \font\fivewedge=stmary5
\def\owedge{%
\def\Dowedge{\mathbin{\hbox{\tenwedge\char"3F}}}%
\def\Sowedge{\mathbin{\hbox{\sevenwedge\char"3F}}}%
\def\SSowedge{\mathbin{\hbox{\fivewedge\char"3F}}}%
\mathchoice{\Dowedge}{\Dowedge}{\Sowedge}{\SSowedge}}

\def\qed{\hbox{\hskip 3pt
\vbox{\hrule\hbox to 7pt{\vrule height 7pt\hfill\vrule}
\hrule}}\hskip3pt}

\overfullrule=0pt\relax

\frenchspacing

\def\checkdef#1#2{
\ifx\UndeFined#1%
	\def#1{#2}
\else
	\immediate\write16{*** BUG ***: the label \string#1 is already defined ***}
\fi
}
\newread\instream
\def\openlabels{
\openin\instream= label.defs
\ifeof\instream\message{No labels in advance yet. Wait till next pass.}
\else\closein\instream \input label.defs
\fi}

\openin\instream= label.defs
\ifeof\instream\message{No labels in advance yet. Wait till next pass.}
\else\closein\instream \input label.defs
\fi
\writedefs

\def\arXiv:#1].{\hepthStrip#1 \nil}
\def\hepthStrip#1 #2\nil{\href{http://arxiv.org/abs/#1}{arXiv:#1 #2\unskip}].}

\font\frakfont=eufm10 at 10pt
\def\ce{\mathord{\hbox{\frakfont e}}}
\def\cf{\mathord{\hbox{\frakfont f}}}
\def\cm{\mathord{\hbox{\frakfont M}}}

\input epsf.tex
\catcode`\@11\relax
\def\youngdim{1.6}
\newif\ify@autoscale \y@autoscaletrue \def\Yautoscale#1{\ifnum #1=0
  \y@autoscalefalse\else\y@autoscaletrue\fi}
\newdimen\y@b@xdim
\newdimen\y@boxdim \y@boxdim=13pt
\def\Yboxdim#1{\y@autoscalefalse\y@boxdim=#1}
\newdimen\y@linethick    \y@linethick=.3pt
\def\Ylinethick#1{\y@linethick=#1}
\newskip\y@interspace \y@interspace=0ex plus 0.3ex
\def\Yinterspace#1{\y@interspace=#1}
\newif\ify@vcenter   \y@vcenterfalse
\def\Yvcentermath#1{\ifnum #1=0 \y@vcenterfalse\else\y@vcentertrue\fi}
\newif\ify@stdtext   \y@stdtextfalse
\def\Ystdtext#1{\ifnum #1=0 \y@stdtextfalse\else\y@stdtexttrue\fi}
\newif\ify@enable@skew   \y@enable@skewfalse
\expandafter\ifx\csname enableskew\endcsname\relax
 \y@enable@skewfalse \else \y@enable@skewtrue\fi
\def\y@vr{\vrule height0.8\y@b@xdim width\y@linethick depth 0.2\y@b@xdim}
\def\y@emptybox{\y@vr\hbox to \y@b@xdim{\hfil}}
\ify@enable@skew
 \def\y@abcbox#1{\if :#1\else
   \y@vr\hbox to \y@b@xdim{\hfil#1\hfil}\fi}
 \def\y@mathabcbox#1{\if :#1\else
   \y@vr\hbox to \y@b@xdim{\hfil$#1$\hfil}\fi}
\else
 \def\y@abcbox#1{\y@vr\hbox to \y@b@xdim{\hfil#1\hfil}}
 \def\y@mathabcbox#1{\y@vr\hbox to \y@b@xdim{\hfil$#1$\hfil}}
\fi
\def\y@setdim{%
  \ify@autoscale%
   \ifvoid1\else\typeout{Package youngtab: box1 not free! Expect an
     error!}\fi%
   \setbox1=\hbox{A}\y@b@xdim=\youngdim\ht1 \setbox1=\hbox{}\box1%
  \else\y@b@xdim=\y@boxdim \advance\y@b@xdim by -2\y@linethick
  \fi}
\newcount\y@counter
\newif\ify@islastarg
\def\y@lastargtest#1,#2 {\if\space #2 \y@islastargtrue
  \else\y@islastargfalse\fi}
\def\y@emptyboxes#1{\y@counter=#1\loop\ifnum\y@counter>0
  \advance\y@counter by -1 \y@emptybox\repeat}
\def\y@nelineemptyboxes#1{%
  \vbox{%
    \hrule height\y@linethick%
    \hbox{\y@emptyboxes{#1}\y@vr}
    \hrule height\y@linethick}\vskip-\y@linethick}
\def\yng(#1){%
  \y@setdim%
  \hskip\y@interspace%
  \ifmmode\ify@vcenter\vcenter\fi\fi{%
  \y@lastargtest#1,
  \vbox{\offinterlineskip
    \ify@islastarg
     \y@nelineemptyboxes{#1}
    \else
     \y@ungempty(#1)
    \fi}}\hskip\y@interspace}
\def\y@ungempty(#1,#2){%
  \y@nelineemptyboxes{#1}
  \y@lastargtest#2,
  \ify@islastarg
   \y@nelineemptyboxes{#2}
  \else
   \y@ungempty(#2)
  \fi}
\def\y@nelettertest#1#2. {\if\space #2 \y@islastargtrue
  \else\y@islastargfalse\fi}
\def\y@abcboxes#1#2.{%
  \ify@stdtext\y@abcbox#1\else\y@mathabcbox#1\fi%
  \y@nelettertest #2.
  \ify@islastarg\unskip%
   \ify@stdtext\y@abcbox{#2}\else\y@mathabcbox{#2}\fi%
  \else\y@abcboxes#2.\fi}
 \newdimen\y@full@b@xdim
 \newcount\y@m@veright@cnt
\ify@enable@skew
 \def\y@get@m@veright@cnt#1#2.{%
   \if :#1 \advance\y@m@veright@cnt by 1\y@get@m@veright@cnt#2.\fi}
 \let\y@setdim@=\y@setdim
 \def\y@setdim{%
   \y@setdim@ \y@full@b@xdim=\y@b@xdim
   \advance\y@full@b@xdim by 1\y@linethick}
 \def\y@m@veright@ifskew#1{
   \y@m@veright@cnt=0 \y@get@m@veright@cnt#1.
   \moveright \y@m@veright@cnt\y@full@b@xdim}
\else
 \def\y@m@veright@ifskew#1{}
\fi
\def\y@nelineabcboxes#1{%
  \y@nelettertest #1.
  \ify@islastarg
   \y@m@veright@ifskew{#1}
    \vbox{
      \hrule height\y@linethick%
      \hbox{\ify@stdtext\y@abcbox#1\else\y@mathabcbox#1\fi\y@vr}
      \hrule height\y@linethick}\vskip-\y@linethick
  \else
   \y@m@veright@ifskew{#1}
    \vbox{
      \hrule height\y@linethick%
      \hbox{\y@abcboxes #1.\y@vr}%
      \hrule height\y@linethick}\vskip-\y@linethick
  \fi}
\def\young(#1){%
  \y@setdim%
  \hskip\y@interspace%
  \y@lastargtest#1,
  \ifmmode\ify@vcenter\vcenter\fi\fi{%
  \vbox{\offinterlineskip
    \ify@islastarg\y@nelineabcboxes{#1}%
    \else\y@ungabc(#1)%
    \fi}}\hskip\y@interspace}
\def\y@ungabc(#1,#2){%
  \y@nelineabcboxes{#1}%
  \y@lastargtest#2,
  \ify@islastarg\y@nelineabcboxes{#2}%
  \else\y@ungabc(#2)%
  \fi}
\catcode`\@12\relax
 

\def\gm{\gamma}
\def\G#1{G^{(#1)}}
\def\B#1{B^{(#1)}}
\def\H#1{H^{(#1)}}
\def\K#1{K^{(#1)}}
\def\C#1{C^{(#1)}}
\def\F#1{F^{(#1)}}

\def\textbf#1{{\bf #1}}
\def\cH{{\cal H}}
\def\paragraph#1{\medskip\noindent{\it #1.}}
\def\paragraphbf#1{\medskip\noindent{\bf #1.}}
\def\rep#1#2{{\bf #1}_{[#2]}}
\def\repr#1[#2]{{\bf #1}_{[#2]}}
\def\et{\widetilde\eta}
\def\so{{\frak so}}
\def\gam{\gamma}
\def\al{\alpha}
\def\be{\beta}
\def\ot{\otimes}
\def\youngdim{0.6}

\title A spectrum-first construction of the second massive open-string vertex

\title in the pure spinor formalism

\author
Chen Huang\email{a}{chen.huang@soton.ac.uk} and
Carlos R. Mafra\email{b}{c.r.mafra@soton.ac.uk}

\address
Mathematical Sciences and STAG Research Centre, University of Southampton,
Highfield, Southampton, SO17 1BJ, UK

\abstract
We construct the unintegrated vertex operator at the
second massive level of the ten-dimensional open
superstring in the pure spinor formalism using a
spectrum-first approach. Exploiting the tensor-product
structure of the massive spectrum, we organize the
physical states into parent and irreducible superfields
and derive their superspace equations of motion.
We identify a system of fundamental massive
superspace equations valid at arbitrary nonzero
mass, from which the parent equations and their
recursive $\theta$ expansions are obtained.
Using the irreducible
superfields and their field strengths as building blocks
for a general vertex ansatz, we solve the BRST-closure
condition and obtain a remarkably compact five-term
vertex. As an application, we compute the three-point
tree-level amplitude involving one second-level massive
state and two massless states and exhibit its relation
to the corresponding amplitude at the first massive
level.
Our spectrum-first construction suggests a systematic
approach to the superspace description of higher
massive levels.

\Date{November 2026}

\lref\fptmass{
	W.~Z.~Feng, D.~Lust, O.~Schlotterer, S.~Stieberger and T.~R.~Taylor,
	``Direct Production of Lightest Regge Resonances,''
	Nucl. Phys. B \textbf{843}, 570-601 (2011)
	[arXiv:1007.5254 [hep-th]].
}

\lref\PSpartfnc{
	Y.~Aisaka, E.A.~Arroyo, N.~Berkovits and N.~Nekrasov,
	``Pure Spinor Partition Function and the Massive Superstring Spectrum,''
	JHEP \textbf{08}, 050 (2008)
	[arXiv:0806.0584 [hep-th]].
}

\lref\lightconeRenann{
	N.~Berkovits and R.~Lipinski Jusinskas,
	``Light-Cone Analysis of the Pure Spinor Formalism for the Superstring,''
	JHEP \textbf{08}, 102 (2014)
	[arXiv:1406.2290 [hep-th]].
}

\lref\DDFRenann{
	R.~L.~Jusinskas,
	``Spectrum generating algebra for the pure spinor superstring,''
	JHEP \textbf{10}, 022 (2014)
	[arXiv:1406.1902 [hep-th]].
}

\lref\FMS{
	D.~Friedan, E.J.~Martinec and S.H.~Shenker,
  	``Conformal Invariance, Supersymmetry and String Theory,''
  	Nucl.\ Phys.\ B {\bf 271} (1986) 93.
}

\lref\Soares{
	B.~R.~Soares,
	``Constructing massive superstring vertex operators from massless vertex operators using the pure spinor formalism,''
	Phys. Lett. B \textbf{852}, 138611 (2024)
	[arXiv:2401.03208 [hep-th]].
}

\lref\massSweden{
       	M.~Guillen, H.~Johansson, R.~L.~Jusinskas and O.~Schlotterer,
	``Scattering Massive String Resonances through Field-Theory Methods,''
	Phys. Rev. Lett. \textbf{127}, no.5, 051601 (2021)
	[\arXiv:2104.03314 [hep-th]].
}

\lref\DDRV{
	G.~D'Appollonio, P.~Di Vecchia, R.~Russo and G.~Veneziano,
	``Microscopic unitary description of tidal excitations in high-energy string-brane collisions,''
	JHEP \textbf{11}, 126 (2013)
	[arXiv:1310.1254 [hep-th]].
}
\lref\FengTaylor{
	W.~Z.~Feng and T.~R.~Taylor,
	``Higher Level String Resonances in Four Dimensions,''
	Nucl. Phys. B \textbf{856}, 247-277 (2012)
	[arXiv:1110.1087 [hep-th]].
}

\lref\olihigher{
	O.~Schlotterer,
	``Higher Spin Scattering in Superstring Theory,''
	Nucl. Phys. B \textbf{849}, 433-460 (2011)
	[arXiv:1011.1235 [hep-th]].
}

\lref\hansenI{
	M.~S.~Costa, T.~Hansen, J.~Penedones and E.~Trevisani,
	``Projectors and seed conformal blocks for traceless mixed-symmetry tensors,''
	JHEP \textbf{07}, 018 (2016)
	[\arXiv:1603.05551 [hep-th]].
}
\lref\hansenII{
	M.~S.~Costa and T.~Hansen,
	``Conformal correlators of mixed-symmetry tensors,''
	JHEP \textbf{02}, 151 (2015)
	[\arXiv:1411.7351 [hep-th]].
}

\lref\cvitanovic{
	P.~Cvitanovic,
	``Group Theory: Birdtracks, Lie's, and Exceptional Groups,''
	Princeton University Press, 2020,
	ISBN 978-0-691-20298-3
}

\lref\fischler{
	M.~Fischler,
	``Young-tableau methods for Kronecker products of representations of the classical groups,''
	J. Math. Phys. 22, 637 (1981)
}
\lref\sagan{
	B.E.~Sagan, ``The symmetric group; representations, combinatorial algorithms and symmetric
	functions'' 2nd edition, Springer, 2001.
}
\lref\gdjames{
	G.D.~James, ``The Representation Theory of the Symmetric Groups'', Springer, 1978.
}
\lref\stahngarnir{
	M.~B.~Green, K.~Peeters and C.~Stahn,
	``Superfield integrals in high dimensions,''
	JHEP \textbf{08}, 093 (2005)
	[\arXiv:hep-th/0506161 [hep-th]].
}

\lref\strathdee{
	J.A.~Strathdee,
	``Extended Poincare supersymmetry,''
	Int. J. Mod. Phys. A \textbf{2}, 273 (1987)
}
\lref\freed{
	D.S.~Freed,
	``Classical Field Theory and Supersymmetry,''
	in Quantum Field Theory: Perspective and Prospective.
}

\lref\ICTP{
	N.~Berkovits,
  	``ICTP lectures on covariant quantization of the superstring,''
	[hep-th/0209059].
}

\lref\weinbergQFTI{
	S.~Weinberg,
	``The Quantum theory of fields. Vol. 1: Foundations,''
	Cambridge University Press, 2005,
	ISBN 978-0-521-67053-1, 978-0-511-25204-4
	doi:10.1017/CBO9781139644167
}
\lref\weinbergQFTIII{
	S.~Weinberg,
	``The quantum theory of fields. Vol. 3: Supersymmetry,''
	Cambridge University Press, 2013,
	ISBN 978-0-521-67055-5, 978-1-139-63263-8, 978-0-521-67055-5
}
\lref\slansky{
	R.~Slansky,
	``Group Theory for Unified Model Building,''
	Phys. Rept. \textbf{79}, 1-128 (1981)
}
\lref\kingBranching{
	R.C.~King,
	``Branching Rules for Classical Lie Groups Using Tensor and Spinor Methods,''
	J. Phys. A \textbf{8}, 429-449 (1975)
}

\lref\PScohomology{
	N.~Berkovits,
	``Cohomology in the pure spinor formalism for the superstring,''
	JHEP \textbf{09}, 046 (2000)
	[arXiv:hep-th/0006003 [hep-th]].
}
\lref\Poincarecohomology{
	N.~Berkovits and O.~Chandia,
	``Lorentz invariance of the pure spinor BRST cohomology for the superstring,''
	Phys. Lett. B \textbf{514}, 394-400 (2001)
	[arXiv:hep-th/0105149 [hep-th]].
}

\lref\bekaert{
	X.~Bekaert and N.~Boulanger,
	``The unitary representations of the Poincar\'e group in any spacetime dimension,''
	SciPost Phys. Lect. Notes \textbf{30}, 1 (2021)
	[arXiv:hep-th/0611263 [hep-th]].
}

\lref\PSthreemass{
	S.~Chakrabarti, S.~P.~Kashyap and M.~Verma,
	``Amplitudes Involving Massive States Using Pure Spinor Formalism,''
	JHEP \textbf{12}, 071 (2018)
	[arXiv:1808.08735 [hep-th]].
}
\lref\oneloopmass{
	C.R.~Mafra,
	``The massive one-loop four-point string amplitude in pure spinor superspace,''
	JHEP \textbf{03}, 145 (2026)
	[arXiv:2508.09930 [hep-th]].
}
\lref\UVshort{
	S.P.~Kashyap, C.R.~Mafra, M.~Verma and L.A.~Ypanaque,
	``A relation between massive and massless string tree amplitudes,''
	[arXiv:2311.12100 [hep-th]].
}

\lref\forcella{
	A.~Hanany, D.~Forcella and J.~Troost,
	``The Covariant perturbative string spectrum,''
	Nucl. Phys. B \textbf{846}, 212-225 (2011)
	[arXiv:1007.2622 [hep-th]].
}
\lref\meka{
	D.~L{\"u}st, N.~Mekareeya, O.~Schlotterer and A.~Thomson,
	``Refined Partition Functions for Open Superstrings with 4, 8 and 16 Supercharges,''
	Nucl. Phys. B \textbf{876}, 55-146 (2013)
	[arXiv:1211.1018 [hep-th]].
}
\lref\LiEprogram{
	M.A.A. van Leeuwen, A.M. Cohen and B. Lisser,
	``LiE, A Package for Lie Group Computations'', Computer Algebra Nederland, Amsterdam, ISBN 90-74116-02-7, 1992
	\semi
	{\tt http://wwwmathlabo.univ-poitiers.fr/\~{}maavl/LiE/}
}
\lref\excursion{
	C.~Markou and E.~Skvortsov,
	``An excursion into the string spectrum,''
	JHEP \textbf{12}, 055 (2023)
	[arXiv:2309.15988 [hep-th]].
}
\lref\deepspectrum{
	T.~Basile and C.~Markou,
	``On the deep superstring spectrum,''
	JHEP \textbf{07}, 184 (2024)
	[arXiv:2405.18467 [hep-th]].
}

\lref\masstheta{
	S.~Chakrabarti, S.~P.~Kashyap and M.~Verma,
	``Theta Expansion of First Massive Vertex Operator in Pure Spinor,''
	JHEP \textbf{01}, 019 (2018)
	[arXiv:1706.01196 [hep-th]].
}

\lref\BCpaper{
	N.~Berkovits and O.~Chandia,
	``Massive superstring vertex operator in D = 10 superspace,''
	JHEP \textbf{08}, 040 (2002)
	[arXiv:hep-th/0204121 [hep-th]].
}
\lref\LieArt{
	R.~Feger, T.~W.~Kephart and R.~J.~Saskowski,
	``LieART 2.0 --- A Mathematica application for Lie Algebras and Representation Theory,''
	Comput. Phys. Commun. \textbf{257}, 107490 (2020)
	[\arXiv:1912.10969 [hep-th]].
}
\lref\threeptm{
	C.~Huang, C.~R.~Mafra and Y.~X.~Tao,
	``Manifest M{\"o}bius invariance of massive tree-level three-point amplitudes in pure spinor superspace,''
	JHEP \textbf{06}, 054 (2026)
	[arXiv:2604.00141 [hep-th]].
}
\lref\bianchispec{
	M.~Bianchi, J.~F.~Morales and H.~Samtleben,
	``On stringy AdS(5) x S**5 and higher spin holography,''
	JHEP \textbf{07}, 062 (2003)
	[arXiv:hep-th/0305052 [hep-th]].
}
\lref\UVpaper{
	S.P.~Kashyap, C.R.~Mafra, M.~Verma and L.~Ypanaqu{\'e},
	``Massless representation of massive superfields and tree amplitudes with the pure spinor formalism,''
	JHEP \textbf{02}, 215 (2025)
	[arXiv:2407.02436 [hep-th]].
}
\lref\treereview{
	C.~R.~Mafra and O.~Schlotterer,
	``Tree-level amplitudes from the pure spinor superstring,''
	Phys. Rept. \textbf{1020}, 1-162 (2023)
	[arXiv:2210.14241 [hep-th]].
}
\lref\FORM{
	J.A.M.~Vermaseren,
	``New features of FORM,''
	arXiv:math-ph/0010025.
\semi
	M.~Tentyukov and J.A.M.~Vermaseren,
	``The multithreaded version of FORM,''
	arXiv:hep-ph/0702279.
}
\lref\towmass{
	C.R.~Mafra,
	``Towards massive field-theory amplitudes from the cohomology of pure spinor superspace,''
	JHEP \textbf{11}, 045 (2024)
	[arXiv:2407.11849 [hep-th]].
}
\lref\intvertex{
	S.~Chakrabarti, S.~P.~Kashyap and M.~Verma,
	``Integrated Massive Vertex Operator in Pure Spinor Formalism,''
	JHEP \textbf{10}, 147 (2018)
	[arXiv:1802.04486 [hep-th]].
}
\lref\psf{
 	N.~Berkovits,
	``Super-Poincare covariant quantization of the superstring,''
	JHEP {\bf 0004}, 018 (2000)
	[arXiv:hep-th/0001035].
}

\lref\wittentwistor{
	E.Witten,
        ``Twistor-Like Transform In Ten-Dimensions''
        Nucl.Phys. B {\bf 266}, 245~(1986)
}

\listtoc
\writetoc
\filbreak

\newsec Introduction

The pure spinor formalism \psf\ provides a manifestly
supersymmetric description of the ten-dimensional
superstring and a framework for the computation of
scattering amplitudes in superspace. While considerable
progress has been made in understanding the interactions
of the massless states of the superstring, the description
of its massive excitations in pure spinor superspace
remains comparatively less developed.

The BRST cohomology of the pure spinor superstring
reproduces the physical spectrum of the theory \refs{\PScohomology,\Poincarecohomology,\PSpartfnc}.
However, the knowledge of the spectrum does not by
itself provide explicit, manifestly Lorentz-covariant superspace representatives
of the corresponding vertex operators. DDF constructions based on light-cone $SO(8)$ superfields
generate physical vertex operators at arbitrary mass levels \refs{\DDFRenann,\lightconeRenann}.
The construction
of covariant representatives requires identifying the
appropriate combinations of worldsheet operators and
the spacetime superfields multiplying them, together
with the superspace constraints that ensure BRST
invariance.

At the first massive level of the open superstring,
the unintegrated vertex operator was constructed by
Berkovits and Chandia \BCpaper, and the corresponding
integrated vertex was subsequently obtained in \intvertex.
The superspace equations of motion and the $\theta$
expansions of the first-massive-level superfields
have been studied in \masstheta, and their applications to
scattering amplitudes have been investigated in
\refs{\PSthreemass,\UVshort,\UVpaper,\towmass,\oneloopmass}.
The first-massive-level vertex also admits a construction from the OPE of massless vertices
\refs{\UVshort,\UVpaper,\Soares,\FMS}.
These developments have made it possible
to describe the interactions of the first massive
supermultiplet in a manifestly supersymmetric manner.
Complementary calculations in the RNS formalism include amplitudes with one first-level massive
state and two or three massless states, with particular attention to four-dimensional
compactifications \fptmass.

In this paper we take a further step in this direction by constructing an explicit unintegrated
vertex operator for the second massive level of the open pure spinor superstring. In addition to
obtaining the vertex itself, we investigate how the superspace description of the second massive
supermultiplet can be related to that of the first massive level. This relation turns out to be
useful both in the construction of the vertex operator and in the computation of scattering
amplitudes.

\newsubsec\specfirstsec A spectrum-first construction of massive vertices

Our strategy for constructing the massive vertex operator differs from that employed by Berkovits
and Chandia \BCpaper\ at the first massive level. In their approach, one starts with the most
general vertex operator compatible with the conformal weight and ghost number of the massive
state. The analysis of its BRST-closure condition and gauge transformations then leads to the
identification of the physical superfields and their superspace equations of motion.

Here we proceed in the opposite direction. We take the representation content of the physical
massive supermultiplet as our starting point and construct a set of covariant superfields
describing its physical degrees of freedom. Having established their superspace equations of
motion, we use these superfields as building blocks for the coefficient superfields of a general
vertex ansatz. The construction of the vertex is then formulated as the problem of finding
combinations of worldsheet operators and physical superfields that satisfy the BRST-closure
condition.

In this way, the problem of identifying the physical supermultiplet and its superspace realization
is separated from the problem of constructing its representative in the BRST cohomology of the
worldsheet theory. We shall refer to this strategy as the {\it spectrum-first approach}. The two
complementary approaches are illustrated in Fig.~1.
A closely related perspective was developed by Basile and Markou, who use representation theory
and Howe duality to construct covariant trajectories of physical states in the NS and Ramond
sectors of the RNS superstring \deepspectrum\foot{Their construction extends the earlier treatment
of the bosonic string by Markou and Skvortsov \excursion.}.

\midinsert
\centerline{\epsfxsize=0.70\hsize
  \epsfbox{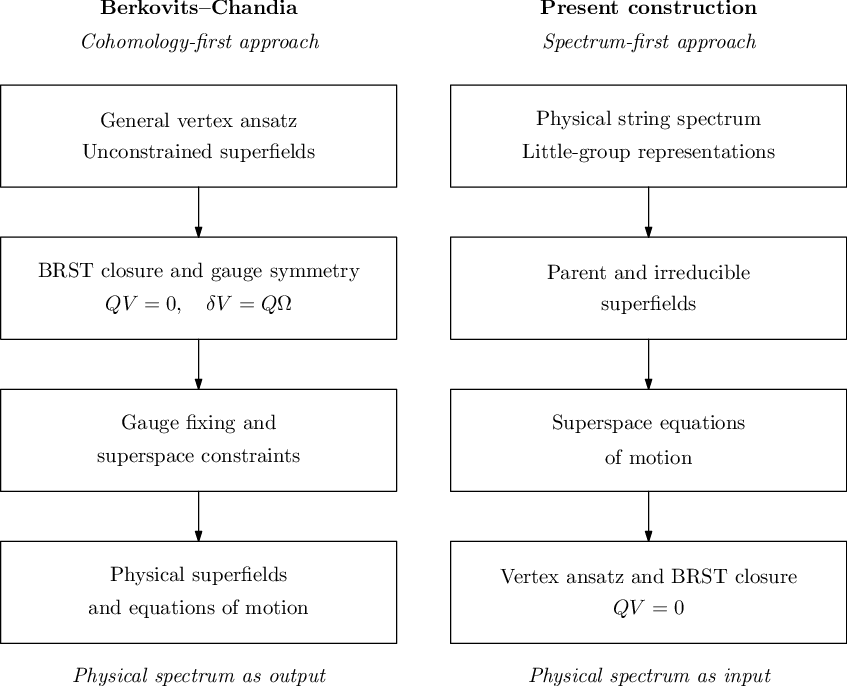}}
\smallskip
{\narrower\noindent
{\bf Fig. 1:} Comparison between the {\it cohomology-first}
approach of Berkovits and Chandia and the
{\it spectrum-first} approach adopted in this work.
\par}
\endinsert

The motivation for adopting this strategy is that the representation content of the massive
supermultiplets is known independently of the explicit construction of their vertex operators. As
the mass level increases, the number of possible worldsheet operator structures grows rapidly,
making a simultaneous determination of the physical superfields, their equations of motion and the
gauge invariances of the vertex increasingly involved. It is therefore advantageous to exploit the
available information about the physical spectrum before addressing the BRST equations.


\newsubsubsec\parirrepsec Parent and irreducible superfields

An important observation underlying our construction
is the organization of the physical massive spectrum
of the open superstring in terms of tensor products
involving a universal massive supermultiplet \forcella.
More specifically, the physical state space ${\cal H}_n$
at a massive level $n$ can be organized as
${\cal H}_n\cong{\cal R}_n\otimes
[(44\oplus84)_B\oplus128_F]$, where ${\cal R}_n$ is a
possibly reducible and graded representation of the
massive little group $SO(9)$, and the representations
inside the square brackets describe what we shall
refer to as the {\it fundamental massive supermultiplet}.
The latter constitutes the entire supermultiplet
at the first massive level of the open superstring.
Its origin in the representation theory of the
massive supersymmetry algebra is explained in
section~\phystatesec.

At the second massive level, ${\cal R}_2$ is the
vector representation of $SO(9)$. Consequently,
the physical spectrum at this level is described
by \forcella
\eqn\HtwoIntro{
 {\cal H}_2\cong 9\otimes
 \big[(44\oplus84)_B\oplus128_F\big]\,,
}
and contains $1152_B+1152_F$ physical degrees of
freedom.

This tensor-product structure suggests introducing
a set of superfields obtained by adjoining an
additional transverse vector index to the three
superfields of the fundamental massive supermultiplet,
\eqn\parentsIntro{
 G_{mn}\,, B_{mnp}\,, H_{m\a}
 \quad\longrightarrow\quad
 G_{r|mn}\,,B_{r|mnp}\,, H_{r|m\a}\,.
}
We refer to the superfields on the left-hand side
of \parentsIntro\ as the {\it fundamental massive superfields}
and to those on the right-hand side as the
{\it parent superfields} of the second massive level.

The decomposition of the corresponding tensor
products into irreducible representations of
$SO(9)$ gives rise to six bosonic
\eqn\bosirrep{
G^{(9)}_m,\; G^{(156)}_{rmn},\;
G^{(231)}_{r|mn},\; B^{(36)}_{mn},\;B^{(126)}_{rmnp},\;B^{(594)}_{r|mnp}
}
and four
fermionic superfields
\eqn\fermirrep{
H^{(16)}_\a,\; H^{(128)\a}_m, \; H^{(432)}_{rm\a},\; H^{(576)}_{rm\a}\,,
}
which we refer to as the
{\it irreducible superfields}. Their lowest
components describe the individual irreducible
physical polarizations of the second massive
supermultiplet.

These two descriptions of the physical
supermultiplet have complementary advantages.
While the irreducible superfields make manifest
the representation content of the physical
spectrum, the parent superfields make manifest
the relation between the supersymmetry
transformations at the first and second massive
levels.

In particular, the additional vector index of
the parent superfields is a spectator under the
action of the superspace covariant derivative.
Upon expressing the mass dependence of the
superspace equations through the transverse
metric \transmetric, the equations of motion of the parent
superfields at the second massive level
\eqnn\parenteomIntro
$$\eqalignno{
D_\a G_{r|mn} &= -{1\over9}\p^p(\g_{pm}H_{r|n})_\a - {1\over9}\p^p(\g_{pn}H_{r|m})_\a\,,
&\parenteomIntro\cr
D_\a B_{r|mnp} &= -{1\over18}\big[\tilde\eta_{ma}\tilde\eta_{nb}(\g^{ab}H_{r|p})_\a + {\rm cyc}(m,n,p)\big],\cr
D_\a H_{r|m\b} &=-{9\over4}G_{r|mn}\g^n_{\a\b}-{3\over2}\p_a B_{r|bcm}\g^{abc}_{\a\b}+{1\over4}\p_a
B_{r|bcd}\g^{mabcd}_{\a\b}\,.
}$$
take the same form as the superspace equations of
the fundamental massive superfields,
\eqnn\massoneeomIntro
$$\eqalignno{
D_\a G^{mn} &= -{1\over9}\p_p(\g^{pm}H^n)_\a - {1\over9}\p_p(\g^{pn}H^m)_\a\,, &\massoneeomIntro\cr
D_\a B_{mnp} &= -{1\over18}\big[\tilde\eta_{ma}\tilde\eta_{nb}(\g^{ab}H_p)_\a + {\rm
cyc}(m,n,p)\big]\,,\cr
D_\a H^m_\b &=-{9\over4}G_{mn}\g^n_{\a\b}-{3\over2}\p_a B_{bcm}\g^{abc}_{\a\b}+{1\over4}\p_a
B_{bcd}\g^{mabcd}_{\a\b}\,,
}$$
The system \massoneeomIntro\ was originally obtained in \masstheta\ in the context of the first
massive level. An important observation is that, when written in terms of the transverse metric,
it reproduces the supersymmetry algebra for arbitrary nonzero mass $M$, provided $k^2=-M^2$. We
shall therefore refer to \massoneeomIntro\ as the {\it fundamental massive superspace equations}.

Together with the spectator nature of the additional vector index, which can be understood from
the representation theory of the massive supersymmetry algebra as explained in
section~\phystatesec, this suggests that the fundamental massive superspace equations may underlie
the superspace description of higher massive levels.

The superspace equations of motion of the ten irreducible superfields can then be obtained from
this simpler system of parent equations by means of the appropriate decomposition maps (see
appendix~\LabelledToMotherapp). Thus, the increased complexity of the irreducible superspace
equations at the second massive level arises largely from the decomposition of the tensor-product
representations, rather than from a more complicated underlying realization of supersymmetry.

The parent formulation also considerably simplifies the computation of $\theta$ expansions. In
fact, the recursive structure of the $\theta$ expansions at the first massive level
\refs{\masstheta,\towmass} is
inherited by the parent superfields at the second massive level. The expansions of the irreducible
superfields \bosirrep\ and \fermirrep\ can subsequently be obtained by applying the decomposition
maps at any desired order in $\theta$.

\newsubsec\VmtwosecIntro The second-massive-level vertex operator

Having established the superspace equations of motion of the physical superfields, we turn to the
construction of the unintegrated vertex operator at the second massive level.

Our starting point is a general ansatz containing $26$ worldsheet operator structures of ghost
number one and conformal weight two at vanishing momentum. The corresponding coefficient
superfields are expressed in terms of the irreducible superfields and their field strengths.

Imposing BRST closure and taking into account the identities among normal-ordered worldsheet
operators leads to a system of equations for the coefficients of the ansatz. The solutions of this
system contain a number of free parameters. A particularly simple choice of these parameters gives
rise to an unintegrated vertex containing only five terms.
Schematically, the resulting vertex has the
structure
\eqn\VmtwoIntro{
 V^{(2)}\sim
 \Pi\partial\theta(\lambda B)
 +\Pi\Pi(\lambda H)
 +\Pi d(\lambda C)
 +\Pi N(\lambda F)
 +\partial\theta\partial\theta(\lambda K)\,,
}
where the coefficient superfields are given
explicitly in section~\Vtwosec\ in terms of the physical
superfields of the second massive level.

Interestingly, the BRST transformations of these coefficient superfields \seceom\ exhibit a close
resemblance to those of the superfields entering the first-massive-level vertex \SYMBRST. This motivates
our choice of notation for the second-level coefficient superfields and provides another
manifestation of the relation between the two massive levels.

The compactness of the resulting vertex makes it particularly convenient for the computation of
scattering amplitudes involving second-level massive states.

\newsubsec\scAsec Massive scattering amplitudes

As a first application of the vertex operator constructed in this paper, we compute the
three-point tree-level amplitude ${\cal A}_{200}$ involving one massive state of the second level
and two massless states.

Using the prescription for massive three-point amplitudes developed in \threeptm, together with
the BRST transformation laws of the coefficient superfields, we obtain a compact expression for
${\cal A}_{200}$ in pure spinor superspace. A particularly interesting feature of the result is
its relation to the corresponding amplitude ${\cal A}_{100}$ involving a massive state of the
first level \UVpaper
\eqnn\parallelIntro
$$\eqalignno{
\cA_{100}(1,2,3) &= -2\ap\langle (\l H^1_m)(\l\g^m W_2)V_3\rangle, &\parallelIntro\cr
\cA_{200}(1,2,3) &= -4i\ap^2\langle (\l H^1_{k^2|m})(\l\g^m W_2)V_3\rangle\,.
}$$
This relation provides an amplitude-level
manifestation of the tensor-product structure
of the second massive supermultiplet \parentsIntro\ and
suggests that the methods developed for the
computation of amplitudes involving first-level
massive states in \towmass\ may admit generalizations to
higher massive levels.

We also evaluate the bosonic components of ${\cal A}_{200}$, illustrating the computational
advantages of the parent formulation. Rather than computing the $\theta$ expansions of the
irreducible superfields directly, we perform the superspace computation using the parent
superfields and decompose the resulting expression into irreducible physical polarizations at the
end of the calculation.

\newnewsec\prelimsec Review and conventions

For reviews of the pure spinor formalism, see \refs{\ICTP,\treereview}.

\newsubsec\vertexsec Vertex operators and equations of motion

The unintegrated and integrated vertex operators describing the massless open string states
are given by \psf
\eqnn\vertices
$$\eqalignno{
V^{(0)}&=\lambda^\alpha A_\alpha\,,&\vertices\cr
U^{(0)}&=\p\theta^\alpha A_\alpha+\Pi^mA_m
+2\alpha'd_\alpha W^\alpha+\alpha'N^{mn}F_{mn}\,.
}$$
The unintegrated vertex at the first massive level is given by \BCpaper
\eqn\VBC{
V(z)= [\p\t^\b \l^\a B_{\a\b}]_0 + [\Pi^m \l^\a H^m_\a]_0 + 2\ap [d_\b C^\b{}_\a\l^\a]_0
+ \ap[N^{mn} [\l^\a F_{\a mn}]_0]_0\,,
}
where the definition of
the normal-ordering bracket $[AB]_0$ is in the appendix A of \UVpaper\ and\foot{We are using the
Schoonschip notation here. See more details below.}
\eqnn\Vonedefs
\eqnn\Fcurlone
$$\displaylines{
B_{\a\b}= \g^{mnp}_{\a\b}B_{mnp},\quad C^\b{}_\a = {1\over4}\p_m(\g^{mnpq})^\b{}_\a
B_{npq},\hfil\Vonedefs\hfilneg\cr
F_{\a mn} = \half\p_m H_{n\a} - \half\p_n H_{m\a} -{i\over16}(\g_m\g^k H_n)_\a
+{i\over16}(\g_n\g^k H_m)_\a\,.\hfil\Fcurlone\hfilneg
}$$
The massless superfields $\big[A_\a, A_m, W^\a, F^{mn}\big]$
and the massive superfields at the first level $\big[ B_{\a\b}, H^m_\a, C^\a{}_\b, F_{\a mn}\big]$
satisfy the equations of motion \refs{\wittentwistor,\BCpaper,\UVpaper}
\eqn\SYMBRST{
\eqalign{
Q A_\a  &= - D_\a V + (\g^m\l)_{\a}A_m\,,\cr
QA_m &= (\l\g^m W) + \p_m V\,,\cr
Q W^\a &= {1\over4}(\l\g^{mn})^\a F_{mn},\cr
Q F_{mn} &=\p_m(\l\g_n W)- \p_n(\l\g_m W)\,,
}\quad
\eqalign{
Q(\l B)_\a &= (\l\g^m)_\a (\l H)_m\,, \cr
Q(\l H^m) &= (\l\g^m C\l)\,,\cr
Q(C\l)^\a &= {1\over4}(\l\g^{mn})^\a (\l F)_{mn}\,,\cr
Q(\l F)_{mn} &= \half\p_m(\l\g_n C\l)- \half\p_n(\l\g_m C\l)\,,
}}
where
\eqn\lfields{
(\l B)_\a = \l^\b B_{\b\a},\quad
(\l H^m)=\l^\a H^m_\a,\quad
(C\l)^\b = C^\b{}_\a \l^\a\,,\quad
(\l F)_{mn} = \l^\a F_{\a mn}\,,
}
and we are using the ``curl'' representative of $F_{\a mn}$. In this representation, the last two terms of
\Fcurlone\ are dropped from the vertex \VBC\ because of the identity
\eqn\ttvan{
[N^{mn}[(\l\g_m\g^k H_n)-(\l\g_n\g^k H_m)]_0]_0=0,
}
due to the normal ordering constraint
$[N^{mn}\l^\b]_0\g^m_{\b\g}  =
\half[J\l^\b]_0\g^n_{\b\g} +
2(\g^n \p\l)_\g$.
In addition, $Q=\l^\a D_\a$ is the BRST charge acting on superfields where
the supersymmetric covariant derivative
$ D_\a={\p\over\p\t^\a}+{i\over2}(\g^m\t)_\a k_m$
satisfies the supersymmetry algebra
\eqn\susyalg{
 \{D_\alpha,D_\beta\}= i k_m \g^m_{\alpha\beta}\,.
}
Furthermore, the transverse metric has rank nine and it is transverse to momenta,
\eqn\transmetric{
 \tilde\eta_{mn}=\eta_{mn}+{k_m k_n\over M^2},\qquad \tilde\eta_{mn}\tilde\eta^{mn}=9,\qquad
 k^m\tilde\eta_{mn} = 0.
}
In a Wick-rotated $SO(10)$, the transverse metric becomes $\tilde\d^{mn}$.
Gamma strings are intrinsically antisymmetrized with unit weight.
Square brackets on tensor
indices denote unaveraged antisymmetrization. Moreover,
we use the mostly plus metric, $k^2=-M^2$ and $\p_m=ik_m$.
At the $n$-th massive level of the open superstring,
the mass is given by $M^2=n/\ap$. We also use the Schoonschip notation, popularized by FORM \FORM,
in which $k_m\g^m$ is written as $\g^k$.

Throughout this paper, we distinguish between the rotation groups $SO(10)$ and $SO(9)$ and their
Lie algebras $\so(10)$ and $\so(9)$. Tensor-product decompositions, Dynkin
labels and branching rules refer to representations of the corresponding Lie algebras. When
spinorial representations are involved, the corresponding group representations are understood to
be those of the double covers $\mathop{\rm Spin}(10)$ and $\mathop{\rm Spin}(9)$.

\newnewsec\phystatesec The physical supermultiplets

\paragraph{Arbitrary mass level}
The structure of the massive supermultiplets can be
understood from the representation theory of the
massive supersymmetry algebra \refs{\strathdee,\weinbergQFTIII}.
At fixed timelike
momentum, the sixteen supersymmetry generators can
be normalized to generate the Clifford algebra
${\rm Cl}_{16}({\bf C})$, whose unique irreducible
module ${\cal S}$ has dimension $256$.
We shall refer to the corresponding supermultiplet
as the {\it fundamental massive supermultiplet}.

Since every finite-dimensional module of this
Clifford algebra is a direct sum of copies of
${\cal S}$, the representation content at each
massive level can be organized in terms of a
possibly reducible and graded multiplicity space
$ {\cal R}_n=({\cal R}_n)_B\oplus({\cal R}_n)_F$ as
\eqn\Sdef{
{\cal H}_n\cong{\cal R}_n\otimes{\cal S}\,.
}
The fundamental massive
supermultiplet ${\cal S}$ decomposes under the massive little
group $SO(9)$ as ${\cal S}=(44\oplus84)_B\oplus128_F$, and therefore
the complete on-shell state space ${\cal H}_n$
is then given by the graded tensor product \forcella
\eqn\Hndef{
 {\cal H}_n\cong{\cal R}_n\otimes
 [(44\oplus84)_B\oplus128_F]\,.
}
In the open superstring, ${\cal S}$
constitutes the entire supermultiplet at the first
massive level \BCpaper.
In this tensor-product realization, the normalized
supersymmetry generators act on ${\cal S}$, leaving
the multiplicity space ${\cal R}_n$ unchanged,
up to the signs associated with the grading of the
tensor product. This observation will be essential
in our construction of the parent superfields below.

With transverse rest-frame indices $i=1,\ldots,9$,
the field representatives of the fundamental
massive supermultiplet are
\eqn\massonereps{
\repr44[2000]:\; G_{(ij)}, \;\d^{ij}G_{ij} = 0,\qquad
\repr84[0010]:\; B_{[ijk]}, \qquad
\repr128[1001]:\; H_{i\a}, \;(\g^{i})^{\a\b}H_{i\b} = 0\,.
}
While the structure \Hndef\ follows from the
representation theory of the massive supersymmetry
algebra, supersymmetry alone does not determine
which multiplicity space ${\cal R}_n$ occurs at
a given mass level. The latter must be extracted
from the string oscillator spectrum, BRST
cohomology, or a refined partition function \refs{\forcella,\meka}.
For the first four massive levels \forcella\
(see also \bianchispec),
\eqn\threeRns{
R_1 = {\bf 1}_B,\quad R_2 = {\bf 9}_B, \quad R_3 = {\bf 44}_B\oplus{\bf 16}_F,\quad
R_4 = {\bf9}_B\oplus{\bf36}_B\oplus{\bf128}_F\oplus{\bf156}_B\,.
}

\newsubsec\massonesec The first massive level

At the first massive level of the open superstring,
${\cal R}_1=1_B$, and hence the physical state space
consists precisely of the fundamental massive
supermultiplet ${\cal S}$. Its realization in terms
of covariant superfields was obtained by Berkovits
and Chandia \BCpaper, and the complete system of
superspace equations of motion was subsequently
obtained in \masstheta. The $128_B\oplus128_F$ states are
described by the Wick-rotated $SO(10)$ superfields
\eqnn\sotenmassonereps
$$\eqalignno{
\repr54[20000]:&\quad{} G_{mn}, \;\;\d^{mn}G_{mn} = 0,\; k^m G_{mn} =0,&\sotenmassonereps\cr
\repr120[00100]:&\quad{} B_{mnp},\;\; k^m B_{mnp} = 0, \cr
\repr144[10010]:&\quad{} H^m_{\a}, \;\;(\g_{m})^{\a\b}H^m_{\b} = 0,\; k_m H^m_\a=0\,.
}$$
These covariant superfields describe the
$so(9)$ irreps $(44\oplus84)_B\oplus128_F$ of the
fundamental massive supermultiplet. The transversality
constraints in \sotenmassonereps\ select the appropriate physical
representations \bekaert\ from the $\so(10)$ to $\so(9)$ branching
rules, as explained in appendix~\branchapp.

At the first massive level, these superfields satisfy
the following system of superspace equations \masstheta,
\eqnn\massoneDG
\eqnn\massoneDB
\eqnn\massoneDH
$$\eqalignno{
D_\a G^{mn} &= -{1\over9}\p_p(\g^{pm}H^n)_\a - {1\over9}\p_p(\g^{pn}H^m)_\a\,, &\massoneDG\cr
D_\a B_{mnp} &= -{1\over18}(\g^{mn}H^p)_\a + {\ap\over18}\p_a\p_m\Big((\g^{an}H^p)_\a
-(\g^{ap}H^n)_\a\Big) + {\rm cyc}(mnp)\,,&\massoneDB\cr
D_\a H^m_\b &=-{9\over4}G_{mn}\g^n_{\a\b}-{3\over2}\p_a B_{bcm}\g^{abc}_{\a\b}+{1\over4}\p_a
B_{bcd}\g^{mabcd}_{\a\b}\,,&\massoneDH
}$$
where
\eqn\Gmndef{
G_{mn} = -{1\over72}\big((D\g_m H_n)+(D\g_n H_m)\big)\,.
}

\newsubsubsec\fundamentalsec The fundamental massive superspace equations

Although the equations \massoneDG--\massoneDH\ were originally
obtained at the first massive level \masstheta, their validity
is not restricted to this particular mass.
To make this property manifest, we rewrite the
coefficient $\ap/18$ in \massoneDB\ as $1/(18M^2)$ and
express the equation in terms of the transverse
metric \transmetric. In this way, equation \massoneDB\ becomes
\eqn\massoneDBex{
D_\a B_{mnp} = -{1\over18}\big[\tilde\eta_{ma}\tilde\eta_{nb}(\g^{ab}H_p)_\a + {\rm cyc}(m,n,p)\big]\,.
}
An important feature of this rewriting is that the
system of equations \massoneDG, \massoneDH\ and \massoneDBex\ is no
longer tied to the mass-shell condition of the first
massive level. Indeed, we have verified explicitly
that these equations reproduce the supersymmetry
algebra \susyalg\ for arbitrary nonzero mass $M$, using
the mass-shell condition $k^2=-M^2$ without specifying
the value of $M$ in terms of $\ap$.

We shall therefore refer to the system \massoneDG, \massoneDH\
and \massoneDBex\ as the {\it fundamental massive superspace
equations}, and to $G_{mn}$, $B_{mnp}$ and $H_{m\a}$
as the {\it fundamental massive superfields}.
The first massive level of the open superstring
corresponds to the particular choice $M^2=1/\ap$.

The validity of the fundamental massive superspace
equations at arbitrary mass, together with the
tensor-product structure \Hndef, provides the basis
for our construction of the parent superfields at
the second massive level.

\newsubsec\masstwosec The second massive level

At the second massive level of the open superstring,
${\cal R}_2={\bf 9}_B$, and hence the physical state space \Hndef\ is
\eqn\secmass{
\cH_2={\bf9}\otimes[({\bf 44}\oplus {\bf 84})_B\oplus {\bf 128}_F]
}
In view of the tensor-product structure \secmass,
we introduce the corresponding $SO(10)$-covariant
parent superfields by adjoining a vector index
to each of the fundamental massive superfields,
\eqn\parentdef{
G_{r|mn},\quad
B_{r|mnp},\quad
H_{r|m\a}\,.
}
The notation for the three parent superfields
reflects their origin in the fundamental massive
superfields \sotenmassonereps. Their representations are given
by the $\so(10)$ tensor products \LiEprogram,
\eqnn\sotentp
$$\eqalignno{
G_{r|mn}:\quad{}\;\; \repr10[10000]\otimes \repr54[20000] &= \repr210'[30000]\oplus\repr320[11000]\oplus\repr10[10000]&\sotentp\cr
B_{r|mnp}:\quad{} \repr10[10000]\otimes \repr120[00100] &= \repr45[01000]\oplus\repr210[00011]\oplus\repr945[10100]\cr
H_{r|m\a}:\quad{} \repr10[10000]\otimes \repr144[10010]
&=\repr16[00010]\oplus\repr144[10001]\oplus\repr560[01010]\oplus\repr720[20010]
}$$
and they inherit the tensorial constraints of
the fundamental massive superfields:
\eqnn\pconstraints
$$\eqalignno{
 G_{r|mn}&=G_{r|nm},\quad \eta^{mn}G_{r|mn}=0,&\pconstraints\cr
 B_{r|mnp}&=-B_{r|nmp}=-B_{r|mpn},\cr
 (\gamma^m)^{\beta\gamma}H_{r|m\gamma}&=0\,.
}$$
The transversality of the original vector indices is kept and
we assume transversality constraints with respect to the new vector index,
\eqnn\transv
$$\eqalignno{
k^mG_{r|mn}&=0,\quad k^mB_{r|mnp}=0,\quad k^mH_{r|m\beta}=0,&\transv\cr
k^rG_{r|mn}&=0,\quad k^rB_{r|mnp}=0,\quad k^rH_{r|m\beta}=0,
}$$
As at the first massive level, the transversality
constraints select from the $\so(10)$ representations
precisely the $\so(9)$ representations describing
the physical degrees of freedom \bekaert. The resulting
representation content is given in \tensupdef\ below,
and the relevant branching rules are discussed
in appendix~\branchapp.

As explained at the beginning of this section,
the supersymmetry generators act on the fundamental
massive supermultiplet in the tensor-product
decomposition \Hndef, leaving the multiplicity
space ${\cal R}_n$ unchanged. At the second massive
level, ${\cal R}_2=9_B$, and hence the additional
vector index $r$ can be chosen to be a spectator
under supersymmetry.

Since the fundamental massive superspace equations
are valid for arbitrary nonzero mass, the parent
superspace equations are obtained by adjoining
the spectator index $r$ to the system \massoneDG,
\massoneDH\ and \massoneDBex,
\eqnn\parenteom
$$\eqalignno{
D_\a G_{r|mn} &= -{1\over9}\p^p(\g_{pm}H_{r|n})_\a - {1\over9}\p^p(\g_{pn}H_{r|m})_\a\,, &\parenteom\cr
D_\a B_{r|mnp} &= -{1\over18}\big[\tilde\eta_{ma}\tilde\eta_{nb}(\g^{ab}H_{r|p})_\a + {\rm cyc}(m,n,p)\big],\cr
D_\a H_{r|m\b} &=-{9\over4}G_{r|mn}\g^n_{\a\b}-{3\over2}\p_a B_{r|bcm}\g^{abc}_{\a\b}+{1\over4}\p_a
B_{r|bcd}\g^{mabcd}_{\a\b}\,.
}$$
The parent equations \parenteom\ inherit the arbitrary-mass
realization of the supersymmetry algebra of the
fundamental massive superspace equations.
Transversality in the additional vector index is
manifestly preserved due to its spectator nature.
The specialization to the second massive level is
achieved by setting $M^2=2/\ap$.

Decomposing the $B_4={\frak so}(9)$ tensor products in the physical description
of the on-shell state space $\cH_2$ in \secmass\ we get \LiEprogram
\eqnn\tensupdef
$$\eqalignno{
\repr9[1000]\otimes\repr44[2000]&={\bf156}_{[3000]}\oplus{\bf231}_{[1100]}\oplus\repr9[1000],&\tensupdef\cr
\repr9[1000]\otimes\repr84[0010]&={\bf36}_{[0100]}\oplus{\bf126}_{[0002]}\oplus{\bf594}_{[1010]},\cr
\repr9[1000]\otimes\repr128[1001]&={\bf16}_{[0001]}\oplus{\bf128}_{[1001]}
                         \oplus{\bf432}_{[0101]}\oplus{\bf576}_{[2001]}.
}$$
The ten irreducible representations appearing in
\tensupdef\ describe the physical polarizations of the
second massive supermultiplet of the open superstring.
We shall refer to the corresponding superfields as
the {\it irreducible superfields}:
\eqn\labeldefs{
 G^{(9)}_m,G^{(156)}_{rmn},G^{(231)}_{r|mn};\qquad
 B^{(36)}_{mn},B^{(126)}_{rmnp},B^{(594)}_{r|mnp};\qquad
 H^{(16)}_\a,H^{(128)\a}_m,H^{(432)}_{rm\a},H^{(576)}_{rm\a},
}
in contrast to the three parent superfields \parentdef. Here and in the following, irreducibility
refers to the representations of the massive little group $SO(9)$, rather than to representations
of the supersymmetry algebra, whose action mixes the ten superfields. Although the irreducible
superfields are written in a manifestly $SO(10)$-covariant notation, their superscripts denote the
dimensions of the corresponding $\so(9)$ representations. In particular, these superscripts count
the independent physical polarizations after imposing the transversality constraints \bekaert,
rather than the dimensions of the corresponding $\so(10)$ representations. The relation between the
two descriptions is explained by the $\so(10)$ to $\so(9)$ branching rules collected in
appendix~\branchapp.
As an illustrative example, we show in detail that the components of an $\so(10)$ tensor-spinor in
the $\repr720[20010]$ representation transform in the $\repr576[2001]$​ representation of $\so(9)$
upon imposing transversality.

The six irreducible superfields in the $G$ and $B$ sectors are bosonic, while the four in the $H$
sector are fermionic. The total number of physical degrees of freedom is that of the second
massive supermultiplet,
\eqn\secdof{
N_B=9+156+231+36+126+594=1152,\qquad
N_F=16+128+432+576=1152.
}
The Young diagrams of the bosonic irreducible superfields are given by\foot{Translating Dynkin labels to
Young diagrams follows \refs{\fischler,\cvitanovic,\hansenII,\oneloopmass}.}
\def\youngdim{0.7}
\eqn\youngGB{
\eqalign{
\G{9}_m &= \yng(1)\cr
\G{156}_{rmn}&=\yng(3)\cr
\G{231}_{r|mn}&=\yng(2,1)
}\qquad
\eqalign{
\B{36}_{mn} &= \yng(1,1)\,,\cr
\B{126}_{rmnp}&=\yng(1,1,1,1)\,,\cr
\B{594}_{r|mnp}&=\yng(2,1,1)
}}
Therefore the non-trivial Garnir relations \refs{\gdjames,\sagan,\stahngarnir,\oneloopmass} are
\eqnn\garnirdefs
$$\eqalignno{
 G^{(231)}_{r|mn}+G^{(231)}_{m|nr}+G^{(231)}_{n|rm}&=0,&\garnirdefs\cr
 B^{(594)}_{r|mnp}-B^{(594)}_{m|rnp}
 +B^{(594)}_{n|rmp}-B^{(594)}_{p|rmn}&=0.
}$$
The $H^{(16)}_\a$, $H^{(432)}_{rm\a}$, and $H^{(576)}_{rm\a}$ fields retain the original
lower chiral spinor index. The vector--spinor instead has its natural
upper index: $H^{(128)}_m{}^\alpha$.
To see this, note that in ten dimensions the upper (Weyl) and lower (anti-Weyl)
spinor nodes of $D_5$ are $[00001]$ and $[00010]$.
The irreps of $r\otimes H_{m\a}$ (used to create the parent
superfield $H_{r|m\a}$) obey, before and after imposing transversality:
\eqnn\otoW
$$\eqalignno{
 {\bf10}_{[10000]}\otimes{\bf144}_{[10010]}
 &={\bf16}_{[00010]}\oplus{\bf144}_{[10001]}\oplus{\bf560}_{[01010]}\oplus{\bf720}_{[20010]}
 &\otoW\cr
\repr9[1000]\otimes\repr128[1001]&={\bf16}_{[0001]}\oplus{\bf128}_{[1001]}
                         \oplus{\bf432}_{[0101]}\oplus{\bf576}_{[2001]}.
}$$
The $\repr144[10001]$ representation on the right-hand side carries a Weyl spinor index,
contrasted with an anti-Weyl spinor index on the left-hand side. It is identified as the
$H^{(128)\a}_m$ field of ${\frak so}(10)$.
Its transverse
gamma-traceless polarization is the ${\bf128}_{[1001]}$ of ${\frak so}(9)$.
This fixes the placement of the ${\frak so}(10)$ spinor index.

The $\repr432[0101]$ is antisymmetric;
the $\repr576[2001]$ is symmetric and metric-traceless. Their gamma traces are
\eqn\gamtrace{
 (\gamma^m)_{\alpha\beta}H^{(128)}_m{}^\beta=0,\qquad
 (\gamma^r)^{\alpha\beta}H^{(432)}_{rm\beta}=0,\qquad
 (\gamma^r)^{\alpha\beta}H^{(576)}_{rm\beta}=0.
}
The equations of motion of the ten irreducible superfields can be obtained from their parent system of
equations and the projections of the appendix~\LabelledToMotherapp. Doing this one gets, for the $G$ sector
\eqnn\DGeoms
$$\eqalignno{
 D_\alpha G^{(9)}_m
& =-{i\over9}(\g^k)_{\alpha\delta}H_m^{(128)\delta}
  -{11i\over81}k^s(\gamma_{sm})_\alpha{}^\beta H^{(16)}_\beta &\DGeoms\cr
 D_\alpha G^{(156)}_{rmn}&=-{2i\over27}k^s\big[
    (\gamma_{sr})_\alpha{}^\beta H^{(576)}_{mn\beta}
   +(\gamma_{sm})_\alpha{}^\beta H^{(576)}_{nr\beta}
   +(\gamma_{sn})_\alpha{}^\beta H^{(576)}_{rm\beta}\big]\cr
 D_\alpha G^{(231)}_{r|mn}
 &=-{i\over27}k^s\big[
   (\gamma_{srm}{}^q)_\alpha{}^\beta
       (H^{(576)}_{nq\beta}-3H^{(432)}_{nq\beta})\cr
 &\qquad\qquad
   +(\gamma_{srn}{}^q)_\alpha{}^\beta
       (H^{(576)}_{mq\beta}-3H^{(432)}_{mq\beta})\big]\cr
 &\quad+{i\over77}k^s\big[
       (\gamma_{srm})_{\alpha\delta}H_n^{(128)\delta}
      +(\gamma_{srn})_{\alpha\delta}H_m^{(128)\delta}\big]\cr
 &\quad-{i\over616}\g^k_{\alpha\delta}\big[
       \et_{rm}H_n^{(128)\delta}+\et_{rn}H_m^{(128)\delta}
          -2\et_{mn}H_r^{(128)\delta}\big].\cr
}$$
And for the $B$ sector
\eqnn\DBeoms
$$\eqalignno{
 D_\alpha B^{(36)}_{mn}
 &=-{1\over9}H^{(432)}_{mn\alpha}
   -{1\over21}\et_m{}^u\et_n{}^v
       (\gamma_{uv}{}^q)_{\alpha\delta}H_q^{(128)\delta} &\DBeoms\cr
& -{7\over216}\et_m{}^u\et_n{}^v
       (\gamma_{uv})_\alpha{}^\beta H^{(16)}_\beta, \cr
 D_\alpha B^{(126)}_{rmnp}
 &=+{1\over72}\et_r{}^u\et_m{}^v\et_n{}^w\et_p{}^t
       (\gamma_{uvwt}{}^{ab})_\alpha{}^\beta H^{(432)}_{ab\beta}\cr
 &\quad+{1\over168}\et_r{}^u\et_m{}^v\et_n{}^w\et_p{}^t
       (\gamma_{uvwt}{}^q)_{\alpha\delta}H_q^{(128)\delta}\cr
 &\quad-{1\over432}\et_r{}^u\et_m{}^v\et_n{}^w\et_p{}^t
       (\gamma_{uvwt})_\alpha{}^\beta H^{(16)}_\beta, \cr
 D_\alpha B^{(594)}_{r|mnp}
 &=+{1\over18}\et_m{}^u\et_n{}^v\et_p{}^w
       (\gamma_{uvw}{}^q)_\alpha{}^\beta
           (H^{(576)}_{rq\beta}+H^{(432)}_{rq\beta})\cr
 &\quad-{1\over72}\et_r{}^u\et_m{}^v\et_n{}^w\et_p{}^t
       (\gamma_{uvwt}{}^{ab})_\alpha{}^\beta H^{(432)}_{ab\beta}\cr
 &\quad+{1\over63}\big[\et_{rm}H^{(432)}_{np\alpha}
               -\et_{rn}H^{(432)}_{mp\alpha}
               +\et_{rp}H^{(432)}_{mn\alpha}\big]\cr
 &\quad+{1\over1848}\et_r{}^u\et_m{}^v\et_n{}^w\et_p{}^t
       (\gamma_{uvwt}{}^q)_{\alpha\delta}H_q^{(128)\delta}\cr
 &\quad-{1\over462}\et_m{}^u\et_n{}^v\et_p{}^w
       (\gamma_{uvw})_{\alpha\delta}H_r^{(128)\delta}\cr
 &\quad+{1\over3234}\big[
          \et_{rm}\et_n{}^u\et_p{}^v(\gamma_{uv}{}^q)_{\alpha\delta}
         -\et_{rn}\et_m{}^u\et_p{}^v(\gamma_{uv}{}^q)_{\alpha\delta}\cr
 &\qquad\qquad
         +\et_{rp}\et_m{}^u\et_n{}^v(\gamma_{uv}{}^q)_{\alpha\delta}
           \big]H_q^{(128)\delta}\,. \cr
}$$
And for the $H$ sector
\eqnn\DHeoms
$$\eqalignno{
 D_\alpha H^{(16)}_\beta
 &=-{9\over4}G^{(9)}_n(\gamma^n)_{\alpha\beta}
   -{3i\over2}k^s B^{(36)}_{mn}(\gamma_s{}^{mn})_{\alpha\beta}
 -{i\over4}k^s B^{(126)}_{mnpq}
                  (\gamma_s{}^{mnpq})_{\alpha\beta},&\DHeoms\cr
 D_\alpha H_m^{(128)\beta}
 &=-{7\over32}\big[8G^{(9)}_m\delta_\alpha{}^\beta
           +\et_m{}^uG^{(9)}_n(\gamma_u{}^n)_\alpha{}^\beta\big]\cr
 &\quad-{9\over4}G^{(231)}_{s|mn}(\gamma^{ns})_\alpha{}^\beta\cr
 &\quad-{11i\over21}k^s\big[7B^{(36)}_{mn}
                (\gamma_s{}^n)_\alpha{}^\beta
        +\et_m{}^uB^{(36)}_{np}(\gamma_{su}{}^{np})_\alpha{}^\beta\big]\cr
 &\quad+{11i\over36}k^t\big[5B^{(126)}_{mnpq}
                (\gamma_t{}^{npq})_\alpha{}^\beta
        +\et_m{}^uB^{(126)}_{npqs}
                (\gamma_{tu}{}^{npqs})_\alpha{}^\beta\big]\cr
 &\quad-{i\over4}k^s B^{(594)}_{m|npq}
                      (\gamma_s{}^{npq})_\alpha{}^\beta,\cr
 D_\alpha H^{(432)}_{rm\beta}
 &=-{9\over448}\et_r{}^a\et_m{}^b
     \bigg[42\et_a{}^u\et_b{}^v\delta_\beta{}^\delta
 -6\et^{uc}\big(
           (\gamma_{ac})_\beta{}^\delta\et_b{}^v
          -(\gamma_{bc})_\beta{}^\delta\et_a{}^v\big)
          -\et^{uc}\et^{vd}
                             (\gamma_{abcd})_\beta{}^\delta\bigg]\cr
 &\qquad\times\bigg[(G^{(231)}_{u|vn}-G^{(231)}_{v|un})
                                   (\gamma^n)_{\alpha\delta}
 +ik^s(B^{(594)}_{u|npv}-B^{(594)}_{v|npu})
                             (\gamma_s{}^{np})_{\alpha\delta}
                +{4i\over7}B^{(36)}_{uv}(\g^k)_{\alpha\delta}\cr
 &\qquad\qquad+2ik^s B^{(126)}_{uvnp}(\gamma_s{}^{np})_{\alpha\delta}
                   \bigg],\cr
 D_\alpha H^{(576)}_{rm\beta}
 &=-{9\over8}\et_r{}^a\et_m{}^b
     \bigg[{9\over11}\et_a{}^u\et_b{}^v\delta_\beta{}^\delta -{1\over11}\et^{uc}\big(
           (\gamma_{ac})_\beta{}^\delta\et_b{}^v
          +(\gamma_{bc})_\beta{}^\delta\et_a{}^v\big)\bigg]\cr
 &\qquad\times\bigg[(2G^{(156)}_{uvn}-G^{(231)}_{n|uv})
                                   (\gamma^n)_{\alpha\delta}
 +ik^s(B^{(594)}_{u|npv}+B^{(594)}_{v|npu})
                             (\gamma_s{}^{np})_{\alpha\delta}\bigg].\cr
}$$
Note that the rank-three boson $\G{156}_{mnp}$ can be obtained
from the covariant derivative of $\H{576}_{mn\a}$,
\eqn\GmnpFromH{
 \G{156}_{mnp}=-{1\over108}\big[
 (D\g_m \H{576}_{np})
 +(D\g_n \H{576}_{pm})
 +(D\g_p\H{576}_{mn})\big].
}
This relation is analogous to the recovery of $G_{mn}$ from $H^m_\a$ at the first massive
level \BCpaper,
\eqn\GmnFromH{
G_{mn} = -{1\over 72}\big( (D\g_m H_n) + (D\g_n H_m)\big)\,.
}
Finally, to summarize the above discussions we list the properties of the irreducible superfields.
\paragraph{Physical constraints} Every vector slot of each irreducible superfield is transverse,
\eqnn\physc
$$\eqalignno{
 k^m\G9_m=0,\quad &k^r\G{156}_{rmn}=0,\quad
 k^r\G{231}_{r|mn}=0,\quad k^m\G{231}_{r|mn}=0,&\physc\cr
 k^m\B{36}_{mn}=0,\quad &k^r\B{126}_{rmnp}=0,\quad
 k^r\B{594}_{r|mnp}=0,\quad k^m\B{594}_{r|mnp}=0,\cr
 k^m\H{128}_m{}^\alpha=0,\quad &k^r\H{432}_{rm\alpha}=0,\quad
 k^r\H{576}_{rm\alpha}=0.
}$$
The symmetries below make all other contractions not shown in \physc\ vanish.
\paragraph{Symmetries} The irreducible superfields satisfy
\eqnn\symlab
$$\eqalignno{
 \G{156}_{rmn}&=\G{156}_{mrn}=\G{156}_{rnm},\quad
 \eta^{rm}\G{156}_{rmn}=0, &\symlab\cr
 \G{231}_{r|mn}&=\G{231}_{r|nm},\quad
 \G{231}_{r|mn}+\G{231}_{m|nr}+\G{231}_{n|rm}=0,\cr
 \eta^{mn}\G{231}_{r|mn}&=0,\quad\eta^{rm}\G{231}_{r|mn}=0,\cr
 \B{36}_{mn}&=-\B{36}_{nm},\cr
 \B{126}_{rmnp}&={1\over24}\B{126}_{[rmnp]},\cr
 \B{594}_{r|mnp}&=-\B{594}_{r|nmp}=-\B{594}_{r|mpn},\cr
 \B{594}_{r|mnp}& -\B{594}_{m|rnp}+\B{594}_{n|rmp}
 -\B{594}_{p|rmn}=0,\quad\eta^{rm}\B{594}_{r|mnp}=0,
}$$
$$\eqalign{
 \H{432}_{rm\alpha}=-\H{432}_{mr\alpha},\quad&
 \H{576}_{rm\alpha}=\H{576}_{mr\alpha},\quad
 \eta^{rm}\H{576}_{rm\alpha}=0,\cr
 (\gamma^m)_{\alpha\beta}\H{128}_m{}^\beta=0,\quad&
 (\gamma^r)^{\alpha\beta}\H{432}_{rm\beta}=0,\quad
 (\gamma^r)^{\alpha\beta}\H{576}_{rm\beta}=0.\cr
}$$

\newsubsec\FSsec Parent and irreducible field strengths

The distinction between parent and irreducible superfields introduced in
section~\masstwosec\ can also be extended to their field strengths.
The various field strengths associated with the ten irreducible superfields
can be obtained from a much smaller set of parent field strengths, whose
construction follows naturally from the tensor-product structure of the
massive supermultiplet.

\proclaim Definition. A {\it field strength} is a first spacetime derivative
of a physical superfield projected onto a specified Young symmetry, with
the appropriate metric and gamma traces removed.

Consider first the fundamental massive superfields $G_{mn}$, $B_{mnp}$
and $H_{m\alpha}$. Their field strengths can be defined as
\eqnn\fundstrengths
$$\eqalignno{
 K_{m|ab}&=i(k_aG_{bm}-k_bG_{am}),&\fundstrengths\cr
 C_{abcd}&=i(k_aB_{bcd}-k_bB_{acd}+k_cB_{abd}-k_dB_{abc}),\cr
 F_{\a ab}&={i\over16}\big[7(k_aH_{b\alpha}-k_bH_{a\alpha})
 -k_s(\gamma_a{}^s H_b)_\alpha+k_s(\gamma_b{}^s H_a)_\alpha\big].\cr
}$$
The first is a traceless hook tensor, the second a four-form, and the
third a gamma-traceless antisymmetric tensor-spinor. Their $\so(10)$
representations are $\repr320[11000]$, $\repr210[00011]$ and
$\repr560[01010]$, respectively. The fermionic strength agrees with the
gamma-completed first-level strength defined in \BCpaper. These constructions
are valid for arbitrary nonzero mass.

\paragraphbf{Parent field strengths}
In analogy with the construction of the parent superspace equations
\parenteom, we define the parent field strengths by adjoining the spectator
vector index $r$ to the fundamental massive field strengths,
\eqnn\parentstrengths
$$\eqalignno{
 {\cal K}_{r|m|ab}&=i(k_aG_{r|bm}-k_bG_{r|am}),&\parentstrengths\cr
 {\cal C}_{r|abcd}&=i(k_aB_{r|bcd}-k_bB_{r|acd}+k_cB_{r|abd}-k_dB_{r|abc}),\cr
 {\cal F}_{r|\a ab}&={i\over16}\big[7(k_aH_{r|b\alpha}-k_bH_{r|a\alpha})
 -k_s(\gamma_a{}^s H_{r|b})_\alpha+k_s(\gamma_b{}^s H_{r|a})_\alpha\big].\cr
}$$
The additional vector index is a spectator and is transverse in all three
strengths. The tensorial symmetries of the fundamental strengths are
preserved in their original indices. In particular,
\eqn\parentstrengthconstraints{
\eqalign{
 {\cal K}_{r|m|ab}+{\cal K}_{r|a|bm}+{\cal K}_{r|b|ma}&=0,\cr
 \eta^{ma}{\cal K}_{r|m|ab}=0,\qquad
 (\gamma^a)^{\alpha\beta}{\cal F}_{r|\b ab}&=0.\cr}}
Unlike the original vector indices of the parent superfields, the curl
indices need not be transverse. The physical constraints imply
\eqn\parentstrengthdiv{
\eqalign{
 k^a{\cal K}_{r|m|ab}&=ik^2G_{r|bm},\cr
 k^a{\cal C}_{r|abcd}&=ik^2B_{r|bcd},\cr
 k^a{\cal F}_{r|\a ab}&={7i\over16}k^2H_{r|b\alpha}.\cr}}
Therefore, at nonzero mass, the three parent superfields can be recovered
from their field strengths by contraction with the momentum. The parent
field strengths do not introduce additional physical degrees of freedom.

\paragraphbf{Decomposition of the parent field strengths}
The ten-dimensional representation content of the three parent field
strengths is described by the tensor products \LiEprogram
\eqn\parentstrengthreps{
\eqalign{
 \repr10[10000]\otimes\repr320[11000]
 &=\repr45[01000]\oplus\repr54[20000]
 \oplus\repr770[02000]\oplus\repr945[10100]
 \oplus\repr1386[21000],\cr
 \repr10[10000]\otimes\repr210[00011]
 &=\repr120[00100]\oplus\repr126[00020]
 \oplus\repr126[00002]\oplus\repr1728[10011],\cr
 \repr10[10000]\otimes\repr560[01010]
 &=\repr144[10010]\oplus\repr560[01001]
 \oplus\repr1200[00110]\oplus\repr3696[11010].\cr
}}
These decompositions provide a natural starting point for understanding
the field strengths associated with the ten irreducible superfields.
However, the decomposition into irreducible $\so(10)$ representations
need not isolate the individual physical polarizations: a given
Lorentz carrier may receive contributions from several irreducible
superfields.
For example, the $\repr1386[21000]$ component of the parent
field strength decomposition in \parentstrengthreps\ also receives contributions from
the physical $\repr9[1000]$ and $\repr231[1100]$ representations. 
This is consistent with the occurrence of all three physical representations of the
$G$ sector ($\bf9$, $\bf156$ and $\bf231$) in the branching of $\repr1386[21000]$ to
$\so(9)$ listed in appendix~\branchapp.
But the field strength $K^{(1386)}_{ab|mn}$ contains only the irreducible $\G{156}_{rmn}$, as discussed in the
section~\FSapp.

To illustrate this distinction, recall the definition of $\G9_b$ in
appendix~\LabelledToMotherapp. The corresponding strength is obtained directly from a
trace of the parent strength,
\eqnn\strengthtraceG
$$\eqalignno{
 \eta^{rm}{\cal K}_{r|m|ab}
 &=i(k_a\eta^{rm}G_{r|mb}
       -k_b\eta^{rm}G_{r|ma}) &\strengthtraceG\cr
 &=i(k_a\G9_b-k_b\G9_a)
 =\K{45}_{ab}.
}$$
Similarly, the cross trace of the parent four-form gives
\eqn\strengthtraceB{
 -\eta^{ra}{\cal C}_{r|abcd}
 =i(k_b\B{36}_{cd}-k_c\B{36}_{bd}+k_d\B{36}_{bc})=\C{120}_{bcd}.}
A further example is the strength of the four-form superfield $\B{126}_{mnpq}$,
which is obtained by complete antisymmetrization,
\eqn\strengthaltB{
-{1\over96}{\cal C}_{[r|abcd]} = C_{rabcd}.
}
The numerical factor follows from our convention of unaveraged
antisymmetrization.

In these examples, the irreducible field strengths
are obtained from the parent field strengths by
means of momentum-independent Lorentz intertwiners.
More generally, the extraction of a particular
physical polarization may require momentum-dependent
projections, as is already the case for the
decomposition maps of appendix~\LabelledToMotherapp.
Explicit extraction maps expressing all the irreducible
field strengths in terms of their parent field strengths
are collected in section~\irreFSapp.

The two constructions of the irreducible field strengths are summarized
in Fig.~2. One may either first decompose the parent superfields and then
construct their strengths, or first construct the parent strengths and
then apply the appropriate covariant extraction maps. Both routes give
the same tensors. The existence of the latter maps follows from
\parentstrengthdiv; explicit formulas are collected in section~\irreFSapp.

\midinsert
\centerline{\epsfxsize=0.70\hsize\epsfbox{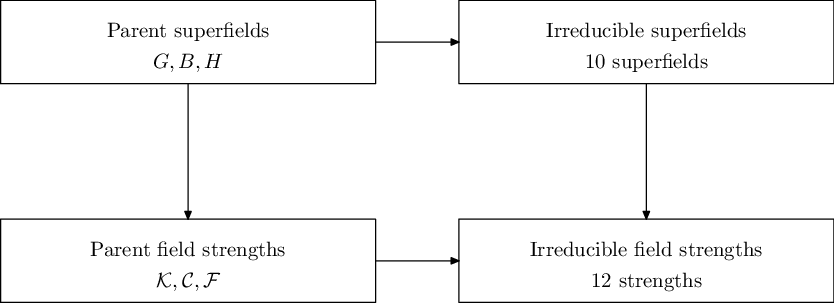}}
\smallskip
{\narrower\noindent{\bf Fig. 2:} Two equivalent constructions of the
irreducible field strengths.\par}
\endinsert

\paragraphbf{Irreducible field strengths}
The ten irreducible superfields give rise to eight bosonic and four
fermionic field-strength tensors. Their representation content is
summarized below. The superscripts of the superfields denote the
dimensions of their physical $\so(9)$ representations, whereas those
of the strengths denote the dimensions of their $\so(10)$ carriers.

\bigskip
{\offinterlineskip
 \halign{\strut\hfil#\hfil&\quad\hfil#\hfil&
          \quad\hfil#\hfil&\quad\hfil#\hfil\cr
 {\bf Physical superfield}&{\bf 10D carrier}&
 {\bf Field strength}&{\bf 10D carrier}\cr
 \noalign{\smallskip\hrule\smallskip}
 $\G9_m$&$\repr10[10000]$&$\K{45}_{ab}$&$\repr45[01000]$\cr
 \noalign{\smallskip\hrule\smallskip}
 $\G{156}_{rmn}$&$\repr210[30000]$&$\K{1386}_{ab|mn}$&$\repr1386[21000]$\cr
 \noalign{\smallskip\hrule\smallskip}
 $\G{231}_{r|mn}$&$\repr320[11000]$&$\K{945}_{r|abc}$&$\repr945[10100]$\cr
 &&$\K{770}_{ab|cd}$&$\repr770[02000]$\cr
 \noalign{\smallskip\hrule\smallskip}
 $\B{36}_{mn}$&$\repr45[01000]$&$\C{120}_{abc}$&$\repr120[00100]$\cr
 \noalign{\smallskip\hrule\smallskip}
 $\B{126}_{rmnp}$&$\repr210[00011]$&$C_{abcde}$&
 $\repr126[00002]\oplus\repr126[00020]$\cr
\noalign{\smallskip\hrule\smallskip}
 $\B{594}_{r|mnp}$&$\repr945[10100]$&$\C{1728}_{r|abcd}$&$\repr1728[10011]$\cr
 &&$\C{2970}_{abc|de}$&$\repr2970[01100]$\cr
\noalign{\smallskip\hrule\smallskip}
 $\H{16}_\alpha$&$\repr16[00010]$&$\F{144}_{m\alpha}$&$\repr144[10010]$\cr
\noalign{\smallskip\hrule\smallskip}
 $\H{128}_m{}^\alpha$&$\repr144[10001]$&$\F{560}_{mn}{}^\alpha$&$\repr560[01001]$\cr
\noalign{\smallskip\hrule\smallskip}
 $\H{432}_{mn\alpha}$&$\repr560[01010]$&$\F{1200}_{mnp\alpha}$&$\repr1200[00110]$\cr
\noalign{\smallskip\hrule\smallskip}
 $\H{576}_{rm\alpha}$&$\repr720[20010]$&$\F{3696}_{r|mn\alpha}$&$\repr3696[11010]$\cr
 \noalign{\smallskip\hrule}
 }}
\bigskip

The field strengths of the bosonic irreducible superfields are defined by
\eqnn\bosFS
$$\eqalignno{
 \C{120}_{abc}&=i (k_a\B{36}_{bc}-k_b\B{36}_{ac}+k_c\B{36}_{ab}),&\bosFS\cr
 C_{abcde}&=i (k_a\B{126}_{bcde}-k_b\B{126}_{acde}+k_c\B{126}_{abde}
             -k_d\B{126}_{abce}+k_e\B{126}_{abcd}),\cr
\C{1728}_{r|abcd}&=i (k_a\B{594}_{r|bcd}-k_b\B{594}_{r|acd}
                        +k_c\B{594}_{r|abd}-k_d\B{594}_{r|abc}),\cr
 \C{2970}_{abc|de}&={i \over3}\big[2(k_d\B{594}_{e|abc}-k_e\B{594}_{d|abc})
 +k_a(\B{594}_{e|bcd}-\B{594}_{d|bce})\cr
 &\qquad-k_b(\B{594}_{e|acd}-\B{594}_{d|ace})
 +k_c(\B{594}_{e|abd}-\B{594}_{d|abe})\big],\cr
 \K{45}_{ab}&=i (k_a\G{9}_b-k_b\G{9}_a),\cr
\K{770}_{ab|cd}&={i \over2}\big[
 k_c(\G{231}_{a|bd}-\G{231}_{b|ad})-k_d(\G{231}_{a|bc}-\G{231}_{b|ac})\cr
 &\qquad+k_a(\G{231}_{c|db}-\G{231}_{d|cb})-k_b(\G{231}_{c|da}-\G{231}_{d|ca})\big].\cr
 \K{945}_{r|abc}&=i \{k_a(\G{231}_{b|cr}-\G{231}_{c|br})
     +k_b(\G{231}_{c|ar}-\G{231}_{a|cr})
     +k_c(\G{231}_{a|br}-\G{231}_{b|ar})\},\cr
\K{1386}_{ab|mn}&=i (k_a\G{156}_{bmn}-k_b\G{156}_{amn}).
}$$
In section~\FSapp\ we explain, as an illustrative example,
why the field strength of $G^{(156)}_{mnp}$ transforms
in the $\bf1386$ representation of $\so(10)$.

The field strengths of the fermionic irreducible superfields are
\eqnn\ferFS
$$\eqalignno{
\F{144}_{m\alpha}&={i\over10}\big[9k_m\H{16}_\alpha
 -k_s(\gamma_m{}^s)_\alpha{}^\delta\H{16}_\delta\big],&\ferFS\cr
\F{560}_{mn}{}^\delta
 &=i  k_{[m}\H{128}_{n]}{}^\delta
   -{i \over8}k_s(\gm_{[m}\g^s)^\delta{}_\sigma
        \H{128}_{n]}{}^\sigma,\cr
 \F{1200}_{mnp\alpha}&={i\over6}\big[5(k_m\H{432}_{np\alpha}
                              +k_n\H{432}_{pm\alpha}+k_p\H{432}_{mn\alpha})\cr
 &\quad-k_s\big((\gamma_m{}^s)_\alpha{}^\delta\H{432}_{np\delta}
               +(\gamma_n{}^s)_\alpha{}^\delta\H{432}_{pm\delta}
               +(\gamma_p{}^s)_\alpha{}^\delta\H{432}_{mn\delta}\big)\big],\cr
 \F{3696}_{r|mn\alpha}&={i\over8}\big[7(k_m\H{576}_{rn\alpha}
                       +k_n\H{576}_{rm\alpha}-2k_r\H{576}_{mn\alpha})\cr
 &\quad-k_s\big((\gamma_m{}^s)_\alpha{}^\delta\H{576}_{rn\delta}
               +(\gamma_n{}^s)_\alpha{}^\delta\H{576}_{rm\delta}
               -2(\gamma_r{}^s)_\alpha{}^\delta\H{576}_{mn\delta}\big)\big].\cr
}$$
The nonzero divergence identities
\eqn\nvdiv{
\eqalign{
 k^m\F{144}_{m\alpha}&={9i\over10}k^2\H{16}_\alpha,\qquad
 k^m\F{560}_{mn}{}^\alpha={7i\over8}k^2\H{128}_n{}^\alpha,\cr
 k^m\F{1200}_{mnp\alpha}&={5i\over6}k^2\H{432}_{np\alpha},\qquad
 k^m\F{3696}_{r|mn\alpha}={7i\over8}k^2\H{576}_{rn\alpha}\cr
}}
recover their respective irreducible superfields at nonzero mass. Therefore, the field strengths do not
introduce new physical data and describe the same degrees of freedom as their associated irreducible
superfields.

Although we have presented the complete set of irreducible
field strengths, only a subset is needed for the construction
of the five-term vertex in section~\Vtwosec. In fact, among the
irreducible field strengths entering the vertex, only
$F^{(560)\a}_{mn}$ requires a momentum-dependent extraction from
its parent field strength. The remaining extraction formulas
in appendix~\irreFSapp\ are included for completeness and may prove
useful in other constructions involving the second massive
supermultiplet.

\newnewsec\Vtwosec A BRST-invariant unintegrated vertex at the second mass level

We now turn to the construction of the unintegrated vertex operator $V^{(2)}$ describing the
physical states at the second massive level of the open superstring. Our starting point is the
most general ansatz compatible with conformal weight two and ghost number one, built from the
worldsheet fields $\Pi^m$, $d_\alpha$, $\partial\theta^\alpha$, $N^{mn}$, $J$, $\l^\a$ and their
derivatives. This ansatz contains 26 operator
structures where each structure is
multiplied by a coefficient superfield constructed from linear combinations of the irreducible
superfields of section~\masstwosec\ and their field strengths, with coefficients to be determined
by BRST invariance.

Requiring BRST invariance,
\eqn\BRSTinv{
QV^{(2)}=0,
}
and taking into account the identities among the worldsheet operators, leads to a coupled system
of superspace equations for the coefficient superfields. We have solved this system and found that
its solutions admit a number of free parameters. Different choices of these parameters lead to
vertex operators with different numbers of terms and different expressions for their coefficient
superfields.

Remarkably, a particularly simple choice of the free parameters yields a vertex operator
containing only five terms. This substantial reduction in the number of operator structures makes
the resulting expression especially convenient for studying the physical superfield content and
computing scattering amplitudes. We shall therefore concentrate on this five-term solution in what
follows.

Although the freedom in the general solution may partly reflect the possibility of adding
BRST-exact terms, we do not need to establish the equivalence of all parameter choices for our
purposes. Instead, we shall demonstrate directly that the five-term vertex is BRST closed and that
its coefficient superfields can be expressed in terms of the irreducible superfields of
section~\masstwosec\ describing the second massive level.

\newsubsec\Vmtwosec An unintegrated vertex at mass level two

The coefficient superfields appearing in the five-term vertex satisfy BRST transformation
equations whose structure closely parallels that of the superfields entering the unintegrated
vertex at the first massive level. In particular, the additional vector index carried by the
second-level superfields suggests an interpretation in terms of vector-valued analogues of the
first-level system.

To make this correspondence manifest, we adopt a
notation modelled on that of the first massive level.
For example, the spinorial coefficient superfield
of $\Pi^m d_\a$ will
be written as $(C_m\lambda)^\alpha$. The five-term unintegrated vertex
is given by
\eqnn\Vmtwo
$$\eqalignno{
V^{(2)} & =
 [\Pi^m[\p\t^\a (B_m\l)_\a]]
+ [\Pi^m[\Pi^n (\l H_{mn})]]
+ (2\ap)[\Pi^m[d_\a (C_m\l)^\a]]&\Vmtwo\cr
&+ \ap[\Pi^m[N^{np}[(\l F_{mnp})]]]
+ [\p\t^\a[\p\t^\b (K\l)_{\a\b}]]
}$$
Its coefficients are
\eqnn\IdfiveAs
\eqnn\IdfiveAsAmn
\eqnn\IdfiveAsW
\eqnn\IdfiveAsF
\eqnn\IdfiveAsK
$$\eqalignno{
(B_m\l)_\b&=-6(\gm^{abc}\l)_{\beta}\B{126}_{mabc}
  +{3\over8}(\gm_m{}^{abcd}\l)_{\beta}\B{126}_{abcd} &\IdfiveAs\cr
 & +{27\over2}(\gm^a\l)_{\beta}\B{36}_{ma}
 +{45\over28}(\gm_m{}^{ab}\l)_{\beta}\B{36}_{ab}
  -{3\over2}(\gm^{abc}\l)_{\beta}\B{594}_{m|abc}\cr
&  +{3 \alpha^\prime\over8}(\gm^{abc}\l)_{\beta}\K{945}_{m|abc}
-{81 \alpha^\prime\over16}(\gm^a\l)_{\beta}\K{45}_{ma}\cr
&  -{27 \alpha^\prime\over64}(\gm_m{}^{ab}\l)_{\beta}\K{45}_{ab}
  -{9i  \alpha^\prime\over14}k_m(\gm^{abc}\l)_{\beta}\C{120}_{abc},\cr
(\l H_{mn}) &=\l^\a\H{576}_{mn\alpha}-{1\over6}\eta_{mn}(\l\H{16})
-{5\over6}\ap k_m k_n (\l\H{16}) &\IdfiveAsAmn\cr
& +{6\over11}[(\l\g_m\H{128}_n)
            +(\l\g_n\H{128}_m)\cr
&-{9\over88}\ap \big[k_m (\l\g^k\H{128}_n) + k_n (\l\g^k\H{128}_m)\big]\cr
 &-{7\over24}\ap\big[k_m(\l\gm_n{}^k\H{16})
                 +k_n(\l\gm_m{}^k\H{16})\big],\cr
(C_m\l)^\rho &= {3\over32}(\g^{abcde}\g_m\l)^\rho {\cal C}_{a|bcde}\,,&\IdfiveAsW\cr
%
(\l F_{mnp})&=(\l\gm_m)_{\delta}{\cal F}_{np}{}^\delta
 -{i \over16}(\l\gm_{mnp})_{\delta}
 k_s(\gm^s)^{\delta\epsilon}\H{16}_\epsilon,&\IdfiveAsF\cr
(K\l)_{\b\g}&=\alpha'(\gm^{abc})_{\beta\gamma}
 \left[-{1\over4}(\l\gm_a)_{\delta}{\cal F}_{bc}{}^\delta
 +{5i \over96}(\l\gm_{abc})_{\delta}
 k_s(\gm^s)^{\delta\epsilon}\H{16}_\epsilon\right].&\IdfiveAsK
}$$
where
\eqn\Fsimp{
 {\cal F}_{mn}{}^\delta={3\over7}\F{560}_{mn}{}^\delta
 -i  k_s(\gm^s)^{\delta\epsilon}\H{432}_{mn\epsilon}.
}
The tensor ${\cal C}_{a|bcde}$ in \IdfiveAsW\ is the full parent
four-form strength defined in \parentstrengths.
From \IdfiveAsW\ it follows that $(\l\g_n C_m\l)$ is symmetric in $(mn)$,
\eqn\Wsym{
(\l\g_m C_n\l) = (\l\g_n C_m\l)\,,
}
and satisfies $(\g^m C_m\l)_\a=0$. Furthermore, one can also show that
\eqn\addids{
\p^p(\l\g_pC_m\l) = -{2\over\ap}(\l B_m\l),\quad
(\l B_m \l) = {3\over8}(\l\g^{mabcd}\l)\B{126}_{abcd}\,.
}
Inspecting \IdfiveAs--\IdfiveAsK\ we see that nine of the ten irreducible superfields appear directly
in the unintegrated vertex \Vmtwo, namely,
\eqn\appear{
 G^{(9)}_m,G^{(231)}_{r|mn};\qquad
 B^{(36)}_{mn},B^{(126)}_{rmnp},B^{(594)}_{r|mnp};\qquad
 H^{(16)}_\a,H^{(128)\a}_m,H^{(432)}_{rm\a},H^{(576)}_{rm\a}\,,
}
the sole exception
being the totally symmetric $\G{156}_{mnp}$. However,
$\G{156}_{mnp}$ appears
in the $\t$-expansion of $\H{576}_{rm\b}$, due to \GmnpFromH,
\eqn\GmnpFromHagain{
 \G{156}_{mnp}=-{1\over108}\big[
 (D\g_m \H{576}_{np})
 +(D\g_n \H{576}_{pm})
 +(D\g_p\H{576}_{mn})\big].
}
therefore the vertex \Vmtwo\ describes all $1152_B+1152_F$ degrees of freedom of the second level massive
supermultiplet. This phenomenon is not new;
at
the first massive level the
totally symmetric $G_{mn}$ superfield is absent from the unintegrated vertex \VBC\ found by Berkovits
and Chandia \BCpaper, which directly contains only $B_{mnp}$ and $H_{m\a}$.
However, the physical degrees of freedom of $G_{mn}$ are accounted by the components of the $\t$
expansion of $H_{m\a}$ due to \GmnFromH.
In this way, the first massive vertex
describes the totality of the $128_B+128_F$ degrees of freedom in the first massive supermultiplet.

Using the equations of motion \DGeoms, \DBeoms\ and \DHeoms\ one can show that the coefficients
of the second level massive vertex \Vmtwo\
satisfy ($\p_m = ik_m$)
\eqnn\seceom
$$\eqalignno{
Q(B_m\l)_\a &=2(\l\g^n)_\a (\l H_{mn}),&\seceom\cr
Q(\l H_{mn}) &= (\l\g_m C_n\l),\cr
Q(C_m\l)^\a &={1\over 4}(\l\g^{np})^\a(\l F_{mnp}),\cr
Q(K\l)_{\a\b}&=\half\big((\l\g^m)_\b (B_m\l)_\a - (\l\g^m)_\a (B_m\l)_\b\big),\cr
Q(\l F_{mnp})&= {1\over2}\big(\p_n(\l\g_p C_m\l) - \p_p(\l\g_n C_m\l)\big)\cr
&+{1\over8\ap}\big((\l\g_m \g_n B_p\l)-(\l\g_m \g_p B_n\l)\big)\cr
&-{1\over45\ap}(\l\g_{mnp}\g^qB_q\l)\cr
&+{5\over36\ap}\big(\d_{mn}(\l B_p\l) - \d_{mp}(\l B_n\l)\big)\,.
}$$
The equation of motion of $(\l F_{mnp})$ is slightly longer in BRST form. Alternatively,
\eqn\DFmnp{
\half \ap (\g^{np}\l)^\a D_\a(\l F_{mnp}) = (\l B_m \l)+ 2\ap \p_n (\l\g^n C_m\l)=-3(\l B_m \l)\,,
}
where we used \addids.
The BRST variation of $V^{(2)}$ yields\foot{The ordering of operators chosen here is
$\big[\l^\a, \p\l^\a, \p^2\l^\a, N^{mn}, \p\t^\a, d_\a, \Pi^m\big]$, where $\l^\a$ is innermost.}
\eqnn\QVtwo
$$\eqalignno{
QV^{(2)} &=
 [\p^2\l^\a Q^{(1)}_\a]
+ [\p\t^\a[\p\l^\b Q^{(2)}_{\a\b}]]
+ [\p\t^\a[\p\t^\b Q^{(3)}_{\a\b}]]
+ [\p\t^\a [N^{mn} Q^{(4)}_{\a mn}]] &\QVtwo\cr
& + [\p^2\t^\a Q^{(5)}_\a]
+ [\Pi^m[\p\l^\a Q^{(6)}_{m\a}]]
+ [\Pi^m[\p\t^\a Q^{(7)}_{m\a}]]
+ [\Pi^m [\Pi^n Q^{(8)}_{mn}]]
+ [\Pi^m [d_\a Q^{(9)\a}_{m}]]\cr
& +[\Pi^m [N^{np} Q^{(10)}_{mnp}]]
+ [d_\a [\p\t^\b Q^{(13)\a}_{\b}]]
}$$
where
\eqnn\Qdefs
$$\eqalignno{
Q^{(1)}_\a &= - \ap (\g^m C_m\l)_\a = 0,&\Qdefs\cr
Q^{(2)}_{\a\b} &=
 - (K\l)_{\a\b}
 + (K\l)_{\b\a}
 - {\ap\over 2}(\l F_{mnp})
   (\g^m\g^{np})_{\a\b}\neq0,\cr
Q^{(3)}_{\a\b} &=
 {1\over 2}\Big[
 - (B_m\l)_\a (\l\g^m)_{\b}
 + (B_m\l)_\b (\l\g^m)_{\a}
 + 2Q(K\l)_{\a\b}
 \Big]  = 0,\cr
Q^{(4)}_{\a mn} &=
\ap (\l F_{pmn})(\l\g^p)_{\a}\neq0,\cr
Q^{(5)}_{\a} &= - {\ap\over 2}(\l F_{mnp})(\g^m\g^{np}\l)_{\a}=0,\cr
Q^{(6)}_{m\a} &=
 2\ap i (C_m\l)^\b(\g^k)_{\a\b}
 + (B_m\l)_\a
 - {\ap\over 2}D_\b(\l F_{mnp})
   (\g^{np})^\b{}_\a \neq 0,\cr
Q^{(7)}_{m\a} &=
 2(\l H_{mn})(\l\g^n)_{\a}
 - Q(B_m\l)_\a = 0,\cr
Q^{(8)}_{mn} &= Q(\l H_{mn})
 - {1\over 2}(\l\g_n C_m\l)
 - {1\over 2}(\l\g_m C_n\l)=0,\cr
Q^{(9)\a}_{m} &=
 - 2\ap Q(C_m\l)^\a
 - {\ap\over 2}(\l F_{mnp})(\g^{np}\l)^\a = 0,\cr
Q^{(10)}_{mnp} &=
 \ap Q(\l F_{mnp})\neq0,\cr
Q^{(13)\a}_{\b} &=
 - 2\ap(C_m\l)^\a(\l\g^m)_{\b}=0.\cr
}$$
Therefore, only $Q^{(2)}$, $Q^{(4)}$, $Q^{(6)}$, $Q^{(10)}$ are not vanishing by themselves. The
others vanish by their equations of motion \seceom\ or due to constraints following from \IdfiveAs.
Hence, the BRST variation becomes,
\eqn\QVtworemains{
QV^{(2)} =
[\p\t^\a[\p\l^\b Q^{(2)}_{\a\b}]]
+ [\p\t^\a [N^{mn} Q^{(4)}_{\a mn}]]
+ [\Pi^m[\p\l^\a Q^{(6)}_{m\a}]]
+ [\Pi^m [N^{np} Q^{(10)}_{mnp}]]=0,
}
which, after using normal ordering constraint identities, can be shown to vanish. More precisely,
after a long calculation one obtains
\eqnn\vanO
$$\eqalignno{
[\p\l^\b Q^{(2)}_{\a\b}] + [N^{mn}[\l^\b\l^\g (Q^{(4)}_{\a mn})_{\b\g}]] &= 0,&\vanO\cr
[\p\l^\a Q^{(6)}_{m\a}] + [N^{np}[\l^\b\l^\g (Q^{(10)}_{mnp})_{\b\g}]] &= 0,
}$$
due to the constraint \BCpaper
\eqn\Nlalaconstraint{
 [N^{st}[\l^\a\l^\b]]
 (\g_s)_{\b\delta}
 -{1\over2}[J[\l^\a\l^\b]]
 (\g^t)_{\b\delta}
={5\over2}\l^\a(\g^t\p\l)_\delta
 -{1\over2}(\g^{st}\l)^\a
 (\g_s\p\l)_\delta.
}

\newnewsec\massampsec The massive three-point tree amplitude $A_{200}$

We are now going to compute the color-ordered tree-level three-point amplitude with one massive vertex of the
second mass level with two massless states. This amplitude is denoted $A_{200}(1,2,3)$. We are
going to use the pure spinor prescription from \psf\ and the calculation will be done in pure spinor
superspace. Therefore, all three-point couplings of the mass-two supermultiplet are presented at the
superspace level.

\newsubsec\psstzzsec Pure spinor superspace derivation

According to \threeptm, if $w_i$ represents the mass level of the $i^{\rm th}$ string being
scattered and if $w_1+w_2-w_3\ge1$ then the tree-level three-point amplitude is given by
\eqn\Awis{
A_{w_1w_2w_3}(1,2,3) = \langle
[[V_1^{(w_1)},V_2^{(w_2)}]_{w_1+w_2-w_3},V_3^{(w_3)}]_{2w_3}\rangle\,.
}
For the $A_{200}$ amplitude it follows that $w_1+w_2-w_3=2$, and therefore
we need the double pole of $V_1^{(2)}$ and $V_2^{(0)}$.
After a short calculation using \Vmtwo\ and the OPE conventions of \UVpaper,
one obtains
\eqn\dop{
[V_1^{(2)}V_2^{(0)}]_2 = -4i \ap^2(C_{k^2}\l)^\a D_\a V_2
-4\ap^2 (\l H_{k^2k^2}) V_2
+ i\ap^2 (\l F_{k^2mn})(\l\g^{mn}A_2)\,.
}
This expression can be rewritten exploiting the cohomology of pure spinor superspace
and the equations of motion \seceom\ as well as the massless SYM single-particle equations.
\eqnn\doubf
$$\eqalignno{
[V_1^{(2)}V_2^{(0)}]_2 &= -4i\ap^2 (\l\g^m C_{k^2}\l)A^2_m -4\ap^2 (\l H_{k^2k^2}) V_2
+ 4i\ap^2 Q\big( (A_2C_{k^2}\l)\big)\,,&\doubf\cr
&= -4i\ap^2 (\l\g^m C_{k^2}\l)A^2_m +4i\ap^2 (\l H_{k^2m}) \p_m V_2 + 4i\ap^2 Q\big( (A_2C_{k^2}\l)\big)\,,\cr
&= -4i\ap^2 (\l\g^m C_{k^2}\l)A^2_m +4i\ap^2 (\l H_{k^2m})QA^m_2 - 4i\ap^2 (\l H_{k^2m})(\l\g^m W_2)\cr
&\qquad{} + 4i\ap^2 Q\big( (A_2C_{k^2}\l)\big)\,,\cr
&= - 4i\ap^2 (\l H_{k^2m})(\l\g^m W_2)
+ 4i\ap^2 Q\big[ (A_2C_{k^2}\l) - (\l H_{k^2m})A^m_2\big]
}$$
To obtain the first line we used the
massless SYM equation $D_\a V_2 = - QA^2_\a + (\l\g^m)_\a A^2_m$ and integrated the
BRST charge by parts using the equations \seceom.
In the second line we used $(\l H_{k^2k^2})V_2=-i(\l H_{k^2m})\p_m V_2$ and $\p_m V_2 = QA^2_m -
(\l\g_m W_2)$ in the third. Finally, we integrated the BRST charge by parts to obtain the fourth
line. Thus the double pole \dop\ is BRST equivalent to the simpler expression
\eqn\dops{
[V_1^{(2)}V_2^{(0)}]_2
\cong - 4i\ap^2 (\l H_{k^2m})(\l\g^m W_2)
}
which is easily checked to be BRST closed using \seceom.
The $A_{200}$ tree amplitude is therefore
\eqn\twozz{
A_{200}(1,2,3) = -4i\ap^2\langle (\l H_{k^2m})(\l\g^m W_2) V_3\rangle\,.
}
Note that $(\lambda H_{mn})$ in \twozz\ is the coefficient
superfield of the unintegrated vertex defined in \IdfiveAsAmn,
and should not be confused with the contraction
$(\lambda H_{m|n})$ of the parent superfield.
Nevertheless, as we shall show below, the amplitude
admits an equivalent expression in terms of the latter,
making its relation to the corresponding amplitude
at the first massive level manifest.

\newsubsec\Relsec Relation between $A_{100}$ and $A_{200}$ of the first and second mass levels

There is a parallel between the pure spinor superspace formulas for the three-point amplitudes
$A_{100}$ and $A_{200}$, where $A_{100}$ is the amplitude involving one massive state of the first
level \UVpaper\ (see also \PSthreemass);
\eqn\fism{
A_{100}(1,2,3) = -2\ap\langle (\l H^1_m)(\l\g^m W_2)V_3\rangle.
}
To see the parallel between the expressions of $A_{100}$ and $A_{200}$, one shows that
\eqn\showz{
\langle \Omega_{k^2m}(\l\g_m W_2)V_3\rangle = 0\,,\qquad
\Omega_{rm}=(\l H_{rm}) - (\l H_{r|m}) = QY_{rm}+(\l\g_m)_\alpha Z_r^\a
}
where
\eqnn\exsol
$$\eqalignno{
Y_{rm}&=9B^{(36)}_{rm}
      +{9i\over8}\ap^2\big(k_mG^{(9)}_r-3k_rG^{(9)}_m\big),&\exsol\cr
 Z_r^\a &=H_r^{(128)\a}-{5\over18}(\g_r H^{(16)})^\a
              -{16\over18}\ap^2 k_r(\g^k H^{(16)})^\a .
}$$
Since $(\l\g^m W_2)V_3$ is BRST closed the first term of $\Omega_{rm}$ decouples from
$\langle \Omega_{k^2m}(\l\g_m W_2)V_3\rangle$,
while
the second drops out due to the pure spinor constraint $(\l\g^m)_\a (\l\g_m)_\b = 0$. Therefore,
\showz\ and \twozz\ imply that
\eqn\secm{
A_{200}(1,2,3) = -4i\ap^2\langle (\l H_{k^2|m})(\l\g^m W_2) V_3\rangle.
}
In summary, the two amplitudes $A_{100}$ and $A_{200}$ have the same structure
\eqnn\parallel
$$\eqalignno{
A_{100}(1,2,3) &= -2\ap\langle (\l H^1_m)(\l\g^m W_2)V_3\rangle, &\parallel\cr
A_{200}(1,2,3) &= -4i\ap^2\langle (\l H^1_{k^2|m})(\l\g^m W_2)V_3\rangle\,.
}$$


\newsubsec\bcomsec Bosonic components

To extract the component amplitudes from the pure
spinor superspace expression \twozz, one inserts the
$\theta$ expansions of the superfields and evaluates
the resulting correlators using the pure spinor measure \psf
\eqn\psmeasure{
 \langle(\l\g^m\theta)(\l\g^n\theta)(\l\g^p\theta)
 (\theta\g_{mnp}\theta)\rangle=1\,.
}
For a review of this procedure, see \treereview.

The expression for $A_{200}$ in terms of the parent
superfield obtained in the previous section is
particularly convenient for the evaluation of its
components. Using the recurrence relations for $\theta$ expansions
of appendix~\thetaapp, we obtain the amplitude directly in
terms of the bosonic polarizations $G_{r|mn}$ and
$B_{r|mnp}$ of the parent superfields\foot{We use the same notation for the bosonic superfields
and their lowest components, which describe the
physical polarizations of the corresponding states.
The distinction should be clear from the context.}.
The result is \FORM
\eqnn\twozzMcomp
$$\displaylines{
A_{200}(1,2,3) = {\ap\over40} B_{k^2|k^2e^2e^3}\hfil\twozzMcomp\hfilneg\cr
       + {i\over160}\Big[ G_{k^2|k^2k^2}(e^2\cdot e^3)\ap^2
       + G_{k^2|k^2e^2} (k^1\cdot e^3)\ap^2
       - G_{k^2|k^2e^3} (k^1\cdot e^2)\ap^2
       + G_{k^2|e^2e^3}\ap \Big].
}$$
Using the inverse decomposition formulas \Gtolabelled\ and
\BtoLabelled\ to express the parent polarizations in terms of
the irreducible physical polarizations, we obtain
\eqn\AtwoLabelled{
A_{200}(1,2,3) =
}
$${} +  {3 \over 7040}\Big[ G^{(9)}_{k^{2}} (k^1 \cdot e^2) (k^1 \cdot e^3) i \ap ^2
 +  G^{(9)}_{k^{2}} (e^2 \cdot e^3) i \ap  \Big]$$
$${} +  {1 \over 160}\Big[ G^{(156)}_{k^{2}k^{2}k^{2}} (e^2 \cdot e^3) i \ap ^2
 +   G^{(156)}_{k^{2}k^{2}e^2} (k^1 \cdot e^3) i \ap ^2
 -   G^{(156)}_{k^{2}k^{2}e^3} (k^1 \cdot e^2) i \ap ^2
 +   G^{(156)}_{k^{2}e^2e^3} i \ap  \Big]$$
$${} +  {1 \over 160}\Big[ G^{(231)}_{k^{2}k^{2}k^{2}} (e^2 \cdot e^3) i \ap ^2
 +   G^{(231)}_{k^{2}k^{2}e^2} (k^1 \cdot e^3) i \ap ^2
 -   G^{(231)}_{k^{2}k^{2}e^3} (k^1 \cdot e^2) i \ap ^2
 +   G^{(231)}_{k^{2}e^2e^3} i \ap  \Big]$$
$${} -  {1 \over 560}\Big[ B^{(36)}_{k^{2}e^2} (k^1 \cdot e^3) \ap
 + B^{(36)}_{k^{2}e^3} (k^1 \cdot e^2) \ap
 - B^{(36)}_{e^2e^3} \Big]$$
$${} +  {1 \over 40}\, B^{(594)}_{k^{2}k^{2}e^2e^3} \ap\,.$$
Note that only five of the six irreducible bosonic
polarizations contribute to this amplitude. The
contribution of $B^{(126)}_{rmnp}$ vanishes because
$B^{(126)}_{k_2 k_2 e_2 e_3}=0$ by antisymmetry.
The coefficient of the $\G{9}_m$ polarization
reflects the normalization of the projection maps in appendix~\LabelledToMotherapp. Furthermore, one can check
using the phase space of the $A_{200}$ amplitude\foot{$k_1\cdot
k_2{=}k_1\cdot k_3{=}1/\ap$, $k_2\cdot k_3 {=}-1/\ap$ and $k_1\cdot k_1{=}-2/\ap$.} that the
amplitude \AtwoLabelled\ is gauge invariant under $\d e^m_{2,3}{=}k^m_{2,3}$..

The polarization $G^{(156)}_{mnp}$ describes the
maximum-spin state at the second massive level, belonging
to the leading Regge trajectory. Its contribution to
the amplitude agrees, up to an overall normalization,
with the $n=2$ specialization of the RNS amplitude
in equation (3.8) of \olihigher, where the couplings of states
on the leading Regge trajectory were studied at arbitrary
mass levels (see also \FengTaylor\ for related amplitudes involving
second-level massive states in four dimensions).

\newnewsec\concsec Conclusions

In this paper we have applied a spectrum-first approach
to construct the unintegrated vertex operator at the
second massive level of the open superstring in the
pure spinor formalism. The resulting BRST-invariant
vertex \Vmtwo\ contains only five terms, with coefficient
superfields built from the ten irreducible superfields
describing the $1152_B+1152_F$ physical degrees of
freedom of the second massive supermultiplet.

An important outcome of our construction is the
identification of the fundamental massive superspace
equations, which reproduce the supersymmetry algebra
for arbitrary nonzero mass. Their tensor-product
extension provides the parent superspace equations
and reveals the common structure underlying the
superspace descriptions of the first and second
massive levels. This relation is also reflected
in the BRST transformations of the vertex coefficient
superfields and in the pure spinor superspace
expressions \parallel\ for the three-point amplitudes
${\cal A}_{100}$ and ${\cal A}_{200}$.

\medskip
\noindent{\bf Future work.}
The validity of the fundamental massive superspace
equations at arbitrary nonzero mass, together with
the tensor-product structure \Hndef\ of the massive
spectrum, suggests that these equations may provide
a common starting point for the superspace
description of all massive levels. It would be
interesting to explore this possibility and
investigate to what extent the spectrum-first
approach can be used to construct the corresponding
vertex operators.

The construction of the integrated vertex operator
$U^{(2)}$ satisfying $QU^{(2)}=\partial V^{(2)}$
is another natural extension of the present work.
Together with the unintegrated vertex constructed
here, it would provide the ingredients for the
computation of more general scattering amplitudes
involving second-level massive states.

It also remains to determine whether the different solutions of the BRST-closure equations are
related by BRST-exact deformations.

Finally, the similarity between the three-point
amplitudes in \parallel\ raises the question of whether
an analogous relation persists at higher multiplicity.
In particular, it would be interesting to investigate
whether the massive field-theory amplitudes involving
a single massive state of the second level admit
simple recursive descriptions available at the first massive level \refs{\massSweden,\towmass}.

We hope to report on some of these questions in
the future.

\bigskip
\noindent{\bf Acknowledgements:} OpenAI's GPT-6 model was used to assist with the calculations.
The authors are responsible for the scientific content.

\appendix{A}{Decomposition of the parent superfields}
\applab\LabelledToMotherapp

\noindent In this appendix we show how to rewrite the ten irreducible
superfields \labeldefs\ in terms of linear combinations of parent superfields \parentdef.

\paragraph{The G sector}
Cross trace and cyclic averaging give
\eqnn\GnineToM
\eqnn\GofsToM
\eqnn\GdtoToM
$$\eqalignno{
 G^{(9)}_m&:=\eta^{rs}G_{r|sm},&\GnineToM\cr
 G^{(156)}_{rmn}
 &:={1\over3}\big(G_{r|mn}+G_{m|nr}+G_{n|rm}\big)
 -{2\over33}\bigl(\tilde{\eta}_{rm}G^{(9)}_n
                 +\tilde{\eta}_{rn}G^{(9)}_m
                 +\tilde{\eta}_{mn}G^{(9)}_r\bigr),&\GofsToM\cr
 G^{(231)}_{r|mn}&:=G_{r|mn}-G^{(156)}_{rmn}
        -{9\over88}\tilde{\eta}_{rm}G^{(9)}_n
        -{9\over88}\tilde{\eta}_{rn}G^{(9)}_m
        +{2\over88}\tilde{\eta}_{mn}G^{(9)}_r.&\GdtoToM
}$$
Thus the inverse is
\eqn\Gtolabelled{
 G_{r|mn}=G^{(156)}_{rmn}+G^{(231)}_{r|mn}
 +{9\over88}\tilde{\eta}_{rm}G^{(9)}_n
        +{9\over88}\tilde{\eta}_{rn}G^{(9)}_m
        -{2\over88}\tilde{\eta}_{mn}G^{(9)}_r.
}
The trace, cyclic average and remaining hook extraction recover the
three fields. These maps act as the identity on their own summands and
annihilate the other two, establishing the direct sum constructively.

\paragraph{The B sector}
The two-form trace, four-form, and hook are
\eqnn\BtsToM
\eqnn\BodsToM
\eqnn\BfnfToM
$$\eqalignno{
B^{(36)}_{mn}&:=\eta^{rs}B_{r|smn},&\BtsToM\cr
B^{(126)}_{rmnp} &:={1\over4}\big(B_{r|mnp}-B_{m|rnp}+B_{n|rmp}-B_{p|rmn}\big),&\BodsToM\cr
B^{(594)}_{r|mnp} &:=B_{r|mnp}-B^{(126)}_{rmnp}
 -{1\over7}\big(\tilde{\eta}_{rm}B^{(36)}_{np}
        -\tilde{\eta}_{rn}B^{(36)}_{mp}
        +\tilde{\eta}_{rp}B^{(36)}_{mn}\big). &\BfnfToM
}$$
Hence
\eqn\BtoLabelled{
 B_{r|mnp}=B^{(126)}_{rmnp}+B^{(594)}_{r|mnp}
 +{1\over7}\big(\tilde{\eta}_{rm}B^{(36)}_{np}
        -\tilde{\eta}_{rn}B^{(36)}_{mp}
        +\tilde{\eta}_{rp}B^{(36)}_{mn}\big).
}
Cross trace recovers $B^{(36)}$, antisymmetrization recovers
$B^{(126)}$, and subtraction recovers $B^{(594)}$.
Their mixed extractions vanish, so this is an actual invertible
direct-sum decomposition.

\paragraph{The H sector}
The natural outer gamma trace has an upper spinor index. Its trace
subtraction and the metric trace define
\eqnn\Hatom
\eqnn\Hbtom
\eqnn\Hctom
\eqnn\Hdtom
$$\eqalignno{
 H^{(16)}_\beta&:=\eta^{rm}H_{r|m\beta},&\Hatom\cr
 H^{(128)}_m{}^\alpha
 &:=(\gamma^r)^{\alpha\beta}H_{r|m\beta}
 -{2\over9}\tilde{\eta}_{mr}(\gamma^r)^{\alpha\beta}H^{(16)}_\beta &\Hbtom\cr
 H^{(432)}_{rm\beta}
 &:=\half\big(H_{r|m\beta}-H_{m|r\beta}\big)
 -{1\over72}\tilde{\eta}_r{}^u\tilde{\eta}_m{}^v
                         (\gamma_{uv})_\beta{}^\delta H^{(16)}_\delta &\Hctom\cr
 &-{1\over14}\bigl[
 \tilde{\eta}_r{}^u(\gamma_u)_{\beta\delta}H^{(128)}_m{}^\delta
 -\tilde{\eta}_m{}^u(\gamma_u)_{\beta\delta}H^{(128)}_r{}^\delta
 \bigr],\cr
 H^{(576)}_{rm\beta}
 &:=\half\big(H_{r|m\beta}+H_{m|r\beta}\big)
 -{1\over9}\tilde{\eta}_{rm}H^{(16)}_\beta&\Hdtom\cr
 &-{1\over22}\bigl[
 \tilde{\eta}_r{}^u(\gamma_u)_{\beta\delta}H^{(128)}_m{}^\delta
 +\tilde{\eta}_m{}^u(\gamma_u)_{\beta\delta}H^{(128)}_r{}^\delta
 \bigr].
}$$
The complete inverse is
\eqnn\HtoIrre
$$\eqalignno{
 H_{r|m\beta}={}&H^{(432)}_{rm\beta}+H^{(576)}_{rm\beta}
 +{1\over9}\tilde{\eta}_{rm}H^{(16)}_\beta
 +{1\over72}\tilde{\eta}_r{}^u\tilde{\eta}_m{}^v
                         (\gamma_{uv})_\beta{}^\delta H^{(16)}_\delta &\HtoIrre\cr
 &+{1\over77}\bigl[
 9\tilde{\eta}_r{}^u(\gamma_u)_{\beta\delta}H^{(128)}_m{}^\delta
 -2\tilde{\eta}_m{}^u(\gamma_u)_{\beta\delta}H^{(128)}_r{}^\delta
 \bigr].
}$$
Metric trace, the gamma trace, and the
two vector symmetries recover all four fields. Each extraction
returns its own field and annihilates the other three; this establishes
the direct sum.

\appendix{B}{Branching rules from ${\frak so}(10)$ to ${\frak so}(9)$ irreps}
\applab\branchapp

\noindent The massive little group \weinbergQFTI\ in ten dimensions is $SO(9)$, so the
physical polarizations of a massive
string state transform in irreducible representations of ${\frak so}(9)$. On the other hand,
the superfields used in the covariant description of the massive vertex operators carry
representations of ${\frak so}(10)$. The relation between these two descriptions is obtained by
restricting the ${\frak so}(10)$ representations to the ${\frak so}(9)$ subalgebra that
leaves the massive momentum invariant.

For a massive momentum satisfying $k^2=-M^2$, we may choose the rest frame $k^m=(M,0,\ldots,0)$.
The transversality constraints then eliminate components carrying longitudinal vector indices,
leaving precisely the representations of the massive little group \bekaert. For example, a symmetric
traceless tensor $G_{mn}$ transforms in the $\repr54[20000]$ of ${\frak so}(10)$, whose restriction is
$\repr54[20000]\longrightarrow\repr44[2000]\oplus\repr9[1000]\oplus\repr1[0000]$.
In the rest frame, the condition $k^mG_{mn}=0$ implies $G_{0i}=0$ for $i=1, \ldots,9$ and $G_{00}=0$; therefore it
removes the $\repr9[1000]\oplus\repr1[0000]$ irreps, leaving the
$\repr44[2000]$ physical
polarizations of the first massive level.

The same reasoning applies to the tensor-spinor representations, with the additional gamma-trace
constraints selecting the appropriate irreducible representations. In particular, the branching
rules provide a useful check that the covariant superfields introduced in section~\masstwosec\ describe the
required physical degrees of freedom at the second massive level.

Below we collect the relevant branching rules \refs{\slansky,\LieArt}, using Dynkin labels $[a_1a_2a_3a_4a_5]$ for
${\frak so}(10)$ and $[b_1b_2b_3b_4]$ for ${\frak so}(9)$.
\eqnn\sotenbranch
$$\eqalignno{
\rep{1}{00000}&\longrightarrow\rep{1}{0000} &\sotenbranch\cr
\rep{10}{10000}&\longrightarrow\rep{9}{1000}\oplus\rep{1}{0000}\cr
\rep{16}{00001}&\longrightarrow\rep{16}{0001}\cr
\rep{16}{00010}&\longrightarrow\rep{16}{0001}\cr
\rep{45}{01000}&\longrightarrow\rep{36}{0100}\oplus\rep{9}{1000}\cr
\rep{54}{20000}&\longrightarrow\rep{44}{2000}\oplus\rep{9}{1000}\oplus\rep{1}{0000}\cr
\rep{120}{00100}&\longrightarrow\rep{84}{0010}\oplus\rep{36}{0100}\cr
\rep{126}{00002}&\longrightarrow\rep{126}{0002}\cr
\rep{126}{00020}&\longrightarrow\rep{126}{0002}\cr
\rep{144}{10001}&\longrightarrow\rep{128}{1001}\oplus\rep{16}{0001}\cr
\rep{144}{10010}&\longrightarrow\rep{128}{1001}\oplus\rep{16}{0001}\cr
\rep{210}{00011}&\longrightarrow\rep{126}{0002}\oplus\rep{84}{0010}\cr
\rep{320}{11000}&\longrightarrow\rep{231}{1100}\oplus\rep{44}{2000}\oplus\rep{36}{0100}\oplus\rep{9}{1000}\cr
\rep{560}{01001}&\longrightarrow\rep{432}{0101}\oplus\rep{128}{1001}\cr
\rep{560}{01010}&\longrightarrow\rep{432}{0101}\oplus\rep{128}{1001}\cr
\rep{720}{20001}&\longrightarrow\rep{576}{2001}\oplus\rep{128}{1001}\oplus\rep{16}{0001}\cr
\rep{720}{20010}&\longrightarrow\rep{576}{2001}\oplus\rep{128}{1001}\oplus\rep{16}{0001}\cr
\rep{770}{02000}&\longrightarrow\rep{495}{0200}\oplus\rep{231}{1100}\oplus\rep{44}{2000}\cr
\rep{945}{10100}&\longrightarrow\rep{594}{1010}\oplus\rep{231}{1100}\oplus\rep{84}{0010}\oplus\rep{36}{0100}\cr
\rep{1050}{10002}&\longrightarrow\rep{924}{1002}\oplus\rep{126}{0002}\cr
\rep{1050}{10020}&\longrightarrow\rep{924}{1002}\oplus\rep{126}{0002}\cr
\rep{1200}{00101}&\longrightarrow\rep{768}{0011}\oplus\rep{432}{0101}\cr
\rep{1200}{00110}&\longrightarrow\rep{768}{0011}\oplus\rep{432}{0101}\cr
\repr1386[21000]&\longrightarrow\repr910[2100]\oplus\repr156[3000]\oplus\repr231[1100]\oplus\repr44[2000]\oplus\repr36[0100]\oplus\repr9[1000]\cr
\rep{1440}{00012}&\longrightarrow\rep{672}{0003}\oplus\rep{768}{0011}\cr
\rep{1728}{10011}&\longrightarrow\rep{924}{1002}\oplus\rep{594}{1010}\oplus\rep{126}{0002}\oplus\rep{84}{0010}\cr
\rep{2970}{01100}&\longrightarrow\rep{1650}{0110}\oplus\rep{495}{0200}\oplus\rep{594}{1010}\oplus\rep{231}{1100}\cr
\rep{3696}{01002}&\longrightarrow\rep{2772}{0102}\oplus\rep{924}{1002}\cr
\rep{3696}{11001}&\longrightarrow\rep{2560}{1101}\oplus\rep{576}{2001}\oplus\rep{432}{0101}\oplus\rep{128}{1001}\cr
\rep{3696}{11010}&\longrightarrow\rep{2560}{1101}\oplus\rep{576}{2001}\oplus\rep{432}{0101}\oplus\rep{128}{1001}\cr
\rep{5940}{01011}&\longrightarrow\rep{2772}{0102}\oplus\rep{1650}{0110}\oplus\rep{924}{1002}\oplus\rep{594}{1010}\cr
\rep{8064}{02001}&\longrightarrow\rep{4928}{0201}\oplus\rep{2560}{1101}\oplus\rep{576}{2001}\cr
}$$
\bigskip
\paragraphbf{Branching versus transversality}
As an illustration of the distinction between branching and imposing
transversality, consider the symmetric tensor--spinor $H_{mn\al}$
transforming in the $\repr720[20010]$ representation of $\so(10)$. It obeys
\eqn\sdzex{
 H_{mn\al}=H_{nm\al},\qquad
 (\gam^m)^{\al\be}H_{mn\be}=0.
}
The tensor product of $\so(10)$ irreps gives $[20000]\otimes[00010]=[10001]\oplus[20010]$
\LiEprogram.
The gamma-trace constraint removes the irrep $[10001]$ and also implies the ordinary trace constraint
$\eta^{mn}H_{mn\al}=0$. This follows by contracting the gamma-trace
with a second gamma matrix and using the Clifford algebra, together
with the symmetry of $H_{mn\al}$.

The ordinary trace constraint is already implied by the
gamma-trace constraint and should therefore not be counted
independently. In $d$ dimensions, a symmetric rank-two tensor
has $d(d+1)/2$ independent components. Tensoring with a
spinor of dimension $N_s$, we obtain a symmetric
tensor--spinor with $N_s d(d+1)/2$ components.
Since the gamma-trace map is surjective\foot{More explicitly, given an arbitrary vector-spinor
$A_n^\alpha$ in $d$ dimensions, one can always construct
a symmetric tensor-spinor $H_{mn\beta}$ such that
$(\g^m)^{\a\b}H_{mn\b}=A_n^\a$;
simply take
$$
H_{mn\alpha}={1\over d+2}\Big[
(\gamma_m)_{\alpha\beta}A_n^\beta
+(\gamma_n)_{\alpha\beta}A_m^\beta
+\eta_{mn}(\gamma^p)_{\alpha\beta}A_p^\beta
\Big]\,.
$$
The Clifford algebra immediately gives the desired result.
}, its image has
dimension $dN_s$. By the rank--nullity theorem, the
dimension of its kernel is
\eqn\dimk{
\dim\ker T_\gamma
=N_s\left({d(d+1)\over2}-d\right)
={N_s d(d-1)\over2}\,.
}
Specializing to $d=10$ and $N_s=16$, we obtain
\eqn\dimHmna{
 \dim H_{mn\al}=55\cdot16-10\cdot16=720.
}
No momentum constraint has been imposed at this stage: the number
$720$ is the dimension of the $[20010]$ irrep of $\so(10)$,
not the number of physical polarizations.

\bigskip
\paragraphbf{Branching to $\so(9)$}
Upon restriction to the massive little group, the representation
branches as \slansky\LieArt
\eqn\sdzbranch{
 \repr720[20010]\longrightarrow
 \repr576[2001]\oplus\repr128[1001]\oplus\repr16[0001].
}
This is a representation-theoretic statement, independent of the
momentum constraint \kingBranching. To see the meaning of this decomposition, choose the
rest frame $k^m=(M,0,\ldots,0)$ and write $m=(0,i)$, with
$i=1,\ldots,9$. The tensor-spinor components are
\eqn\tscomp{
 H_{ij\al},\qquad H_{0i\al},\qquad H_{00\al}.
}
With all spinor indices displayed, the ten-dimensional gamma-trace
equations become
\eqn\gamtraces{
 (\gam^0)^{\al\be}H_{0j\be}
 +(\gam^i)^{\al\be}H_{ij\be}=0,
 \qquad
 (\gam^0)^{\al\be}H_{00\be}
 +(\gam^i)^{\al\be}H_{0i\be}=0.
}
The mixed component $H_{0i\al}$ is a nine-dimensional vector-spinor.
It decomposes into a gamma-traceless vector-spinor $\psi_{i\al}$ and a
spinor $\chi^\be$ of opposite chirality:
\eqn\Hzialdec{
 H_{0i\al}=\psi_{i\al}+(\gam_i)_{\al\be}\chi^\be,
 \qquad (\gam^i)^{\al\be}\psi_{i\be}=0,
 \qquad \repr9[1000]\ot\repr16[0001]=\repr128[1001]\oplus\repr16[0001].
}
Using $(\gam^i)^{\al\be}(\gam_i)_{\be\gamma}
=9\delta^\al{}_\gamma$, the second gamma-trace equation in \gamtraces\ implies
\eqn\gtrimp{
 (\gam^0)^{\al\be}H_{00\be}+9\chi^\al=0,
 \qquad \chi^\al=-{1\over9}(\gam^0)^{\al\be}H_{00\be},
}
and hence \Hzialdec\ becomes
\eqn\Hzialbec{
 H_{0i\al}=\psi_{i\al}
 -{1\over9}(\gam_i)_{\al\be}(\gam^0)^{\be\gamma}
 H_{00\gamma}.
}
The positions of the spinor indices exhibit the chirality change
under each gamma matrix. Thus the
longitudinal components are parametrized by a $128$-component
$\psi_{i\al}$ and a $16$-component $H_{00\al}$. The spinor part of $H_{0i\al}$
is not an additional independent $16$: it is fixed by $H_{00\al}$.

The first equation in \gamtraces\ determines the gamma trace
of $H_{ij\a}$ in terms of $H_{0j\a}$.
Its remaining independent part is therefore the symmetric
gamma-traceless tensor-spinor of $\so(9)$.
Specializing the general counting formula \dimk\
to $d=9$ and $N_s=16$, we obtain
\eqn\kerdim{
\dim\ker T_\gamma={16\cdot9\cdot8\over2}=576\,.
}
The restriction of the irrep $\repr720[20010]$ of $\so(10)$ to irreps of $\so(9)$
is therefore accounted for by the
independent transverse gamma-traceless part $\repr576[2001]$ and the
longitudinal data $\repr128[1001]\oplus\repr16[0001]$:
\eqn\bransdz{
 \repr720[20010]=\repr576[2001]\oplus\repr128[1001]\oplus\repr16[0001]\,,
}
which explains the corresponding entry of the data in \sotenbranch.

\bigskip
\noindent{\bf Imposing transversality.}\enspace
If the physical polarization additionally obeys
\eqn\traH{
 k^mH_{mn\al}=0,
}
we conclude,
in the rest frame $k^m=(M,0,\ldots,0)$ with $M\ne0$, that $H_{0n\al}=0$. Or in other words,
\eqn\krestc{
H_{00\al}=H_{0i\al}=0.
}
The explicit decomposition \Hzialbec\ then shows that $\psi_{i\al}=0$ and
$\chi^\al=0$: therefore removing the $\repr128[1001]\oplus\repr16[0001]$ irreps.
The remaining gamma-trace constraint is
$(\gam^i)^{\al\be}H_{ij\be}=0$,
which also implies $\delta^{ij}H_{ij\al}=0$. The surviving $H_{ij\al}$ is
the symmetric gamma-traceless rank-two tensor-spinor of $\so(9)$ transforming as
$\repr576[2001]$.

In summary, the $\so(10)$ representation branches as
\eqn\brsum{
\repr720[20010]\longrightarrow\repr576[2001]\oplus\repr128[1001]\oplus\repr16[0001],
}
while the transversality constraint $k^mH_{mn\al}=0$ selects the $\so(9)$ irrep
\eqn\sels{
\repr720[20010]\longrightarrow\repr576[2001].
}
The first arrow denotes restriction of a representation from
$\so(10)$ to $\so(9)$; the second selects the physical polarization
subspace by the momentum constraint. In particular, branching itself
does not eliminate components.

\appendix{C}{Theta expansions at the second massive level}
\applab\thetaapp

\noindent In this appendix we describe the computation of
the $\theta$ expansions of the superfields at the second
massive level. We first explain the advantages of working
with the parent superfields and then derive recurrence
relations for their $\theta$ expansions from the
fundamental massive superspace equations.

\newsubsec\parentthetasec Theta expansions and the parent superfields

At the second massive level, the ten irreducible superfields
\labeldefs\ and the three parent superfields \parentdef\ provide
equivalent descriptions of the physical supermultiplet.
Although the former make the irreducible representation
content of the supermultiplet manifest, the latter are
particularly convenient for the computation of $\theta$
expansions. Indeed, the $\theta$ expansions of the ten
irreducible superfields can be obtained from those of the
three parent superfields by means of the decomposition
formulas collected in appendix~\LabelledToMotherapp.

To see this, let ${\cal P}_R$ denote the linear map
that extracts from a parent superfield $K$ the irreducible
superfield $K^{(R)}$ associated with the irreducible
representation $R$ of the massive little group,
\eqn\calPdef{
 K^{(R)}={\cal P}_R K\,.
}
Since the decomposition maps are constructed out of
momenta, invariant tensors and gamma matrices, they
are independent of $\theta$. Therefore, denoting by
$[K]_n$ the homogeneous component of degree $n$ in
$\theta$ of the superfield $K$, we have
\eqn\calPonKn{
 [K^{(R)}]_n={\cal P}_R[K]_n\,.
}
In particular, the same decomposition formulas that
relate the parent and irreducible superfields also relate
their $\theta$-independent polarizations.

For the purpose of evaluating pure spinor superspace
expressions in components, this observation suggests
a more economical procedure in which the decomposition
into irreducible physical polarizations is postponed
until the end of the calculation. Namely, one first
rewrites the irreducible superfields in terms of their
parents and evaluates the pure spinor superspace
correlators using the $\theta$ expansions of the latter.
The resulting expression, written in terms of the
parent polarizations and momenta, is then expressed
in terms of the irreducible physical polarizations
using the inverse decomposition formulas of appendix~\LabelledToMotherapp.
This is the procedure adopted in section~\bcomsec\ to obtain
the bosonic components of the three-point amplitude
${\cal A}_{200}$.

The advantage of working with the parent superfields
$G_{r|mn}$, $B_{r|mnp}$ and $H_{r|m\a}$ follows
from the fact that their equations of motion \parenteom\
are a tensor-product extension of the fundamental
massive superspace equations. Since the additional
vector index $r$ is a spectator under the action
of the superspace covariant derivative, the recursive
structure of the $\theta$ expansions is inherited
from that of the fundamental massive superfields.

\newsubsec\recsec The recurrence relations for the theta expansion

The recurrence relations for the $\theta$ expansions of the fundamental massive superfields were
originally obtained in \masstheta\ in the context of the first massive level (see also \towmass).
Their extension to the parent superfields follows directly from the parent superspace equations
\parenteom, as we shall now demonstrate.

The contraction of the superspace covariant derivative
with $\theta^\a$ reduces to the operator that counts
the degree in $\theta$,
\eqn\eulerop{
 \t^\a D_\a
 =\t^\a{\p\over\p\t^\a}\,,
 \qquad
 \t^\a D_\a[K]_n=n[K]_n\,,
}
since the term involving the spacetime derivative
vanishes by virtue of $(\theta\g^m\theta)=0$.
Contracting the equations of motion \parenteom\ of the
parent superfields with $\theta^\a$ from the left
then yields the following recurrence relations:
\eqnn\eomthetarec
$$\eqalignno{
[G_{r|mn}]_k &= -{1\over 9k}\Big[ik_p(\t\g^{pm}[H_{r|n}]_{k-1}) + ik_p(\t\g^{pn}[H_{r|m}]_{k-1})\Big] &\eomthetarec\cr
[B_{r|mnp}]_k &= -{1\over18k}\Big[
\tilde\eta_{ma}\tilde\eta_{nb}(\t\g^{ab}[H_{r|p}]_{k-1})\Big] + {\rm cyc}(m,n,p),\cr
[H_{r|m_\b}]_{k} &=-{1\over 4k}\Big[9[G_{r|mn}]_{k-1}(\t\g^n)_{\b}
+6ik_a [B_{r|bcm}]_{k-1}(\t\g^{abc})_{\b}
-ik_a[B_{r|bcd}]_{k-1}(\t\g^{mabcd})_{\b}\Big]\,.
}$$
starting with
\eqn\startrec{
[G_{r|mn}]_0 = G_{r|mn}\,\quad
[B_{r|mnp}]_0 = B_{r|mnp}\,,\quad
[H_{r|m\a}]_0 = \psi_{r|m\a}\,.
}
The recurrence relations \eomthetarec, together with the initial conditions \startrec, determine
the $\theta$ expansions of the three parent superfields to any desired order. In practice, one
starts with the $\theta$-independent polarizations and uses \eomthetarec\ successively to obtain
the homogeneous components of increasing degree in $\theta$. Note that the $\theta$ expansions of
the bosonic parent superfields are determined at each order by the fermionic parent superfield at
the preceding order, and vice versa.

The $\theta$ expansions of the ten irreducible superfields are subsequently obtained by applying
the decomposition formulas of appendix~\LabelledToMotherapp\ at each order in $\theta$, as follows
from \calPonKn. For the sake of brevity, we refrain from displaying the explicit $\theta$
expansions, since all their coefficients are determined by the recurrence relations \eomthetarec\
and the initial conditions \startrec.

\appendix{D}{Construction and decomposition of the field strengths}
\applab\FSappNew

\noindent In this appendix we elaborate on the construction and decomposition of the field strengths
introduced in section~\FSsec. We first illustrate how the Lorentz representation of a field
strength is determined by considering the example of $\G{156}_{mnp}$. We then derive explicit
expressions for the irreducible field strengths in terms of their parent field strengths.

\newsubsec\FSapp The field strength of $\G{156}_{mnp}$

\noindent The field $G^{(156)}_{bmn}$ is symmetric, traceless
and transverse:
\eqn\apsym{
 \eta^{bm}G^{(156)}_{bmn}=0,\qquad k^bG^{(156)}_{bmn}=0.
}
The symmetric traceless rank-three tensor in $d$ dimensions has
dimension ${d+2\choose 3}{-}d$.
This gives 156 physical polarizations
for $d=9$ and a Lorentz carrier of dimension 210 for $d=10$.

\def\youngdim{0.6}
Set $T_{a|bmn}=ik_aG^{(156)}_{bmn}$. Adding the derivative index
to the three symmetric field indices gives the Young diagrams $\yng(4)$
and $\yng(3,1)$. The curl
\eqn\Kabdef{
 K_{ab|mn}=T_{a|bmn}-T_{b|amn}
}
annihilates the completely symmetric part $\yng(4)$. It is antisymmetric
in $a,b$, symmetric in $m,n$, and satisfies the Garnir identity
\refs{\gdjames,\sagan,\stahngarnir,\oneloopmass}
associated with the Young diagram
$\yng(3,1)$
\eqn\garn{
 K_{ab|mn}+K_{bm|an}+K_{ma|bn}=0.
}
The coefficient of $ik_a$ in this sum is
$G^{(156)}_{bmn}-G^{(156)}_{mbn}=0$. The coefficients of $ik_b$
and $ik_m$ cancel by the same symmetry.

A trace over two field indices vanishes by tracelessness, and a (cross) trace
between the derivative index and a field index vanishes by
transversality. For example,
\eqn\exKab{
 \eta^{am}K_{ab|mn}
 =i\bigl(k^mG^{(156)}_{bmn}-k_b\eta^{am}G^{(156)}_{amn}\bigr)=0.
}
The ten-dimensional Lorentz decomposition is \LiEprogram
\eqn\tendL{
 \repr10[10000]\otimes\repr210[30000]=\repr660[40000]\oplus\repr1386[21000]\oplus\repr54[20000].
}
The $\bf660$ is the symmetric traceless rank-four
representation with diagram $\yng(4)$, the $\bf1386$ is the traceless $\yng(3,1)$
carrier, and the $\bf54$ is the symmetric traceless
rank-two representation $\yng(2)$ associated with the
cross trace of $T_{a|bmn}$.
The tracelessness of $G^{(156)}_{bmn}$ is already
incorporated into its $\bf210$ representation, while
its transversality implies that the cross trace
vanishes,
\eqn\crosstr{
 \eta^{ab}T_{a|bmn}=ik^bG^{(156)}_{bmn}=0\,.
}
This removes the $\bf54$ component from the tensor
product \tendL. The curl \Kabdef\ subsequently
annihilates the completely symmetric $\bf660$
component, leaving the field strength $K_{ab|mn}$
in the $\bf1386$ representation of $\so(10)$.
The curl here starts with the symmetric irreducible superfield. The same
Lorentz carrier in the parent strength also receives contributions from
the physical $9$ and $231$ sectors.

Contracting with the momentum recovers the original field:
\eqn\reco{
G^{(156)}_{bmn}={k^a\over ik^2}K^{(1386)}_{ab|mn}.
}

\newsubsec\irreFSapp Extraction of the irreducible field strengths

The divergence identities \parentstrengthdiv\ allow us to recover the
parent superfields from their field strengths. Together with the
maps of appendix A and the definitions \bosFS\ and \ferFS, they
provide a construction of all the irreducible field strengths.
Below we give explicit formulas for their extraction.

We denote by ${\rm TF}_{\lambda}$ the projection
onto the completely traceless part of a tensor
with Young symmetry $\lambda$. The tensors on
which these projectors act are understood to
have already been projected onto the corresponding
Young symmetry. Explicit projectors onto traceless tensors of prescribed
Young symmetry are discussed in \hansenI.

\paragraphbf{The $G$ sector}
In addition to \strengthtraceG, the momentum-independent maps are
\eqn\Gstrengthmaps{\eqalign{
 \K{945}_{r|abc}&={\rm TF}_{[10100]}
 ({\cal K}_{b|r|ac}+{\cal K}_{c|r|ba}+{\cal K}_{a|r|cb}),\cr
 \K{770}_{ab|cd}&={1\over2}{\rm TF}_{[02000]}
 \big({\cal K}_{a|b|cd}-{\cal K}_{b|a|cd}
 +{\cal K}_{c|d|ab}-{\cal K}_{d|c|ab}\big).\cr}}
Here ${\rm TF}_{[10100]}$ and ${\rm TF}_{[02000]}$ remove the
metric traces of tensors with Young symmetries $(2,1,1)$ and $(2,2)$,
respectively \refs{\fischler,\cvitanovic,\oneloopmass}. The arguments of these maps already have the indicated
Young symmetries. For example, for a tensor $T_{r|abc}$ antisymmetric
in $abc$ and satisfying $T_{[r|abc]}=0$, one has
\eqn\TFuzu{
 {\rm TF}_{[10100]}(T)_{r|abc}=T_{r|abc}
 -{1\over8}(\eta_{ra}T^s{}_{|sbc}
 -\eta_{rb}T^s{}_{|sac}+\eta_{rc}T^s{}_{|sab}).
}
Contracting with $\eta^{ra}$ gives zero, since the contraction of the
three metric terms gives $(10-2)T^s{}_{|sbc}$.
Similarly, for a tensor $T_{ab|cd}$ with the symmetries of the Riemann
tensor,
\eqnn\TFzdz
$$\eqalignno{
 {\rm TF}_{[02000]}(T)_{ab|cd}&=T_{ab|cd}
 -{1\over8}(\eta_{ac}T^s{}_{b|sd}-\eta_{ad}T^s{}_{b|sc}
-\eta_{bc}T^s{}_{a|sd}+\eta_{bd}T^s{}_{a|sc})\cr
 &\qquad+{1\over72}(\eta_{ac}\eta_{bd}-\eta_{ad}\eta_{bc})
 T^{st}{}_{|st}. &\TFzdz
}$$
The remaining strength is obtained using the transverse metric,
\eqnn\Gstrengthsym
$$\eqalignno{
 \K{1386}_{ab|mn}&=\sum_{\rm cyc(b,m,n)}\left(
 {1\over3}\et_n{}^v{\cal K}_{b|m|av}
 +{2\over33}\et_{bm}\eta^{rs}{\cal K}_{r|n|sa}\right)-(a\leftrightarrow b)&\Gstrengthsym\cr
&=ik_a\G{156}_{bmn}-ik_b\G{156}_{amn}\,.
}$$

\paragraphbf{The $B$ sector}
The strengths $\C{120}_{abc}$ and $C_{abcde}$ are given by
\strengthtraceB\ and \strengthaltB, respectively. 
Taking the curl of the inverse map \BtoLabelled\ gives
\eqn\Bstrengthhook{\eqalign{
 \C{1728}_{r|abcd}&={\cal C}_{r|abcd}+C_{rabcd}
 -{k_r k^s\over k^2}C_{sabcd}\cr
 &\quad+{1\over7}(\eta_{ra}\C{120}_{bcd}-\eta_{rb}\C{120}_{acd}
 +\eta_{rc}\C{120}_{abd}-\eta_{rd}\C{120}_{abc}).\cr}}
The other strength of $\B{594}_{r|abc}$ can then be written as
\eqn\Bstrengthother{
 \C{2970}_{abc|de}={k_d k^s\over k^2}\C{1728}_{e|sabc}
 -{k_e k^s\over k^2}\C{1728}_{d|sabc}
 -{1\over3}(\C{1728}_{e|dabc}-\C{1728}_{d|eabc}).
}
Substituting the definition of $\C{1728}$ in \bosFS\ reproduces
the expression for $\C{2970}$ in the same equation.

\paragraphbf{The $H$ sector}
The trace of the parent strength determines the $16$ and $128$
 superfields through
\eqnn\Hstrengthtrace
$$\eqalignno{
 \H{16}_\a&=-{16\over7ik^2}k^b\eta^{ra}{\cal F}_{r|\a ab},&\Hstrengthtrace\cr
 \H{128}_b{}^\a &={16\over ik^2}(\g^s)^{\a\b}k_s
 \left(\eta^{ra}{\cal F}_{r|\beta ab}
 +{35\over72}\F{144}_{b\beta}\right).
}$$
Their field strengths follow from \ferFS. In particular the identity
underlying these maps is
\eqn\idun{
 \eta^{ra}{\cal F}_{r|\alpha ab}
 ={i\over16}k_s(\gamma^s)_{\alpha\beta}\H{128}_b{}^\beta
 -{35\over72}\F{144}_{b\alpha}.
}
The strength of $\H{432}$ is obtained by antisymmetrizing the
vector indices of the parent strength and removing the gamma traces,
\eqn\Hstrengthalt{
 \F{1200}_{abc\alpha}=-{\rm GT}_{[00110]}
 \big({\cal F}_{a|\alpha bc}+{\cal F}_{b|\alpha ca}
 +{\cal F}_{c|\alpha ab}\big).}
For an antisymmetric tensor-spinor $T_{abc\alpha}$ the gamma-trace
subtraction in this formula is explicitly
\eqnn\exform
$$\eqalignno{
 {\rm GT}_{[00110]}(T)_{abc\alpha}&=T_{abc\alpha}
 -{1\over6}\sum_{\rm cyc(a,b,c)}(\gamma_a\gamma^d)_\alpha{}^\beta
 T_{dbc\beta}&\exform\cr
 &\quad+{1\over42}\sum_{\rm cyc(a,b,c)}
 (\gamma_{ab}\gamma^d\gamma^e)_\alpha{}^\beta T_{edc\beta}\cr
 &\quad-{1\over336}(\gamma_{abc}\gamma^d\gamma^e\gamma^f)_\alpha{}^\beta
 T_{fed\beta}.
}$$
Finally, the symmetric tensor-spinor is obtained from the divergence
of the parent strength by subtracting the other three physical
components according to the inverse map  \HtoIrre,
\eqn\Hstrengthsym{\eqalign{
 \H{576}_{rm\alpha}&={16k^a\over7ik^2}{\cal F}_{r|\alpha am}
 -\H{432}_{rm\alpha}-{1\over9}\et_{rm}\H{16}_\alpha\cr
 &\quad-{1\over72}\et_r{}^a\et_m{}^b
 (\gamma_{ab})_\alpha{}^\beta\H{16}_\beta\cr
 &\quad-{1\over77}(9\et_r{}^a\gamma_{a\alpha\beta}\H{128}_m{}^\beta
 -2\et_m{}^a\gamma_{a\alpha\beta}\H{128}_r{}^\beta).\cr}}
Here $\H{432}_{rm\alpha}=6k^a\F{1200}_{arm\alpha}/(5ik^2)$ by
\nvdiv. The expression for $\F{3696}_{r|mn\alpha}$ is then obtained
from \ferFS.

\ninepoint
\listrefs

\bye